\documentclass[onecol]{aa}

\usepackage{xcolor}
\usepackage{epstopdf}
\usepackage{pdflscape}
\def\oiiiopt{{\sc{[Oiii]}}$\lambda\lambda$4959,5007\/}
\def\oii{{\sc{[Oii]}}$\lambda$3727\/}
\def\nii{{\sc{[Nii]}}$\lambda$6583\/}
\def\sii{{\sc{[Sii]}}$\lambda\lambda$6716,6731\/}
\def\oi{{\sc{[Oi]}}$\lambda$6300\/}

\def\fe{{\sc{Fe}}\/}

\def\fe76087{{\sc [Fe vii]}$\lambda$6087\/}
\def\oiii{{\sc [Oiii]}$\lambda$5007}

\def\hb{H$\beta$\/}
\def\ha{H$\alpha$\/}
\def\kms{km~s$^{-1}$}

\def\ergss{erg s$^{-1}$\/}
\def\hii{H{\sc ii}\/}
\def\hi{H{\sc i}\/}

\def\mvabs{$M_\mathrm{V}$\/}
\def\re{$R_\mathrm{e}$\/}
\def\msol{$M_\odot$\/}
\def\ddnii{DD$_\mathrm{[NII]}$\/}
\def\ddoii{DD$_\mathrm{[OII]}$\/}
\def\ddsii{DD$_\mathrm{[SII]}$\/}
\def\ddoi{DD$_\mathrm{[OI]}$\/}
\def\cst{w0$_\mathrm{t,M}$\/}
\def\csres{$\overline{\mathrm{\mathrm w0}}_\mathrm{t,M,R}$\/}

\begin{document}
\authorrunning{Marziani, D' Onofrio, Bettoni, Poggianti, et al.}
\titlerunning{AGNs in WINGS}
\title{Emission Line Galaxies and Active Galactic Nuclei \\ in WINGS clusters}
 \author{P. Marziani\inst{1}, M. D'Onofrio\inst{1,2}, D. Bettoni\inst{1},   B. M. Poggianti\inst{1},  A. Moretti\inst{1}, \\ G. Fasano\inst{1},  
  J. Fritz\inst{3}, A. Cava,$^{4}$\ J. Varela$^{5}$, A. Omizzolo\inst{1,6}}
\institute{INAF, Osservatorio Astronomico di Padova, vicolo dell' Osservatorio 5, IT 35122, Padova, Italy. 
\email{paola.marziani@oapd.inaf.it}
\and Dipartimento di Fisica \& Astronomia ``Galileo Galilei'', Universit\`a di Padova, vicolo dell' Osservatorio 3, IT 35122, Padova, Italy 
\and Centro de Radioastronom\'\i a y Astrof\'\i sica, CRyA, UNAM, Michoac\'an, Mexico
\and Observatoire de Gen\`eve, Universit\'e de Gen\`eve, 51 Ch. des Maillettes, 1290 Versoix, Switzerland
\and Centro de Estudios de F\'\i sica del Cosmos de Arag\'on (CEFCA), Plaza San Juan 1, planta 2, 44001 Teruel, Spain
\and Specola Vaticana, 00120, Vatican City State}
\date{}
\abstract
{{  We present the analysis of the emission line galaxies members of 46 low redshift ($0.04<z<0.07$) clusters observed by WINGS (WIde-field Nearby Galaxy cluster Survey, \citealt{fasanoetal06}).} Emission line galaxies were identified following criteria that are meant to minimize biases against non-star forming galaxies and classified employing diagnostic diagrams. We have examined the emission line properties and  frequencies of star forming galaxies, transition objects and active galactic nuclei (AGNs: LINERs and Seyferts), unclassified  galaxies with  emission lines, and quiescent galaxies with no detectable line emission.  
A deficit of emission line galaxies in the cluster environment is indicated by  both a lower frequency  with respect to control samples, and by a systematically   lower Balmer emission line equivalent width and luminosity (up to one order  of magnitude in equivalent width with respect to control samples for transition objects) that implies a lower amount of ionised gas per unit mass and a lower star formation rate if the source is classified as \hii\ region.  A sizable  population of transition objects  and of low-luminosity LINERs ($\approx 10 - 20 $\% of all emission line galaxies)  is detected among WINGS cluster galaxies. With respect to   \hii\ sources they are a factor of $\approx$ 1.5 more frequent than  (or at least as frequent as)  in control samples.  Transition objects and LINERs in cluster are    most affected in terms of {  line equivalent} width by the  environment and appear predominantly consistent with   ``retired'' galaxies.  Shock heating can be a possible gas excitation  mechanism able to account for  observed line ratios. Specific to the cluster environment, we suggest interaction between atomic and molecular gas and the intracluster medium as a possible physical cause of line-emitting shocks.
}
\maketitle


\section{Introduction}
{ Many} galaxy properties depend on their environment (see e.g. \citealt{sulentic76,larsontinsley78}  for pionieering work and e.g.\ \citealt{leeetal03,poggiantietal06,bernardietal06,blantonmoustakas09,scovilleetal13}, {  \citealt{bitsakisetal16}}\ for more modern perspectives). At one extreme we find the most isolated galaxies whose morphology may be a fossil of early times of galaxy formation \citep{verdes-montenegroetal05}. At the other extreme we find galaxies living in the very dense environment of clusters, where morphology and gas content are affected by frequent inter-galaxy gravitational interactions \ such as minor and major merging, various forms of tidal harassments involving {\sc HI} disk truncation,  and tidal stripping of gas and stars \citep[e.g.,][]{daleetal01,parkhwang09}. If we were considering only gravitational effects, a na\"{\i}ve expectation would be to find a large fraction of luminous emission line galaxies (ELGs) hosting an active nucleus right among cluster galaxies: strong gravitational interactions are known to provide a trigger for nuclear activity over a wide range of luminosity, if one of the galaxies involved in the interaction is sufficiently gas rich  \citep{hwangetal12,sabateretal13}.  It has been recognized since over 30 years  that this is not the case of  cluster environment, most likely because of the interaction of the galaxy atomic and molecular gas with  the  hot intra-cluster medium (ICM) whose pressure is higher in  the inner cluster core ($\lesssim  0.5 r_\mathrm{vir}$).  
The dynamical action of the ICM  on a galaxy interstellar gas may lead to its   stripping by ram  pressure. Ram stripping is expected to   influence the    evolution of galaxies in clusters to the point of even transforming  a spiral galaxy into an ``anaemic'' S0 galaxy \citep[e.g.,][]{spitzerbaade51,gunngott72,giovanellihaynes85,boselligavazzi06,donofrioetal15}.   
Systematic  effects of ram stripping have been traced by a deficit of  {\sc Hi} gas,   spectacularly illustrated for the Virgo cluster \citep{cayatteetal90}, and   by a deficit of ELGs in cluster environment \citep[e.g.,][]{balickheckman82,hwangetal12,pimbbletetal13}.

{Of particular interest in this context is the problem of the AGN fraction and location in galaxy clusters \citep[e.g.,][]{milleretal03,Kauffmannetal2004,Choietal2009,martinietal06,martinietal09,haggardetal2010,lindenetal2010,hwangetal12,pimbbletetal13}. These works often ended with contrasting results concerning the role of environment even when the extended spectral database provided by the SDSS
is used. It is possible that the origin of these discrepancies is linked to the employed data samples  as well as to  analysis technique adopted for the emission line spectra.
It is also possible that 
two major open issues concerning ELGs and their properties plague any conclusion about the AGN fraction and the role of the cluster environment: 1) the classification of galaxies as possible AGNs by means of diagnostic diagrams (DDs) might vary with the DD used and strongly depends on the signal-to-noise (S/N) ratio of the available spectra, i.e. considering the errors; 2) the understanding of the physical nature of faint ELGs as possible AGNs is far from definitive.}


Faint emission lines are buried in the absorption spectrum of the host galaxy, as spectra are often obtained with a fibre covering a significant spatial extent of the host galaxies.  DDs are customarily used for the classification of ELGs. They however require  the accurate measurement of 3 or 4 emission lines. The nowadays most-frequently used DDs are the ones proposed by \citet{veilleuxosterbrock87}. They have the considerable advantage of employing line pairs that are proximate in wavelength. The backside is that 2 of 3 DDs involve lines that are fairly weak, and they require a measure of the \ha\ and \hb\ emission lines that are most affected by the absorption spectrum of the host galaxy. 


The second  issue is related to the physical nature of ELGs. Once a source  is placed in a DD, the customary subdivision involves a net separation of sources in three classes: {\sc Hii}, LINERs and Seyfert sources, the latter two often collectively referred to as AGNs. In the DD  involving \nii/\ha\ vs \oiii/\hb\ a finer subdivision has been introduced. A line based on the theoretical maximum line ratios possible from pure stellar photoionization distinguishes \hii\ regions from other ELGs \citep{kewleyetal01}. Sources above this line (i.e., with larger \oiii/\hb\ ratio) are likely dominated by emission  associated with non-stellar sources of ionization as provided by  active nuclei. The \citet{kewleyetal01} dividing line comes from  a photoionization analysis that includes extreme condition for line emission from gas ionized by stellar sources. A further subdivision    \citep{kauffmannetal03} distinguishes between    pure star-forming galaxies and  Seyfert and ``composite'' objects or ``transition'' objects (TOs) i.e., sources located in between the region of LINERs and of \hii\ region in the DD \oiii/\hb\ vs \nii/\ha. This latter line is purely empirical.   While classification is safe and not questioned for most \hii\ regions, the physical interpretation of the DDs involves two controversial aspects: (1)   Seyferts (with high ionization spectra) and LINERs (by definition with low ionization spectra) are included in the broader class of AGNs; (2) TOs  show properties that may not be fully consistent with photoionization by hot stars but still lie within the region of theoretically admissible \hii\ region. 

Most galaxies in the field show emission lines, if the equivalent width is as low as 0.25 \AA. Their line luminosity is correspondingly  low ($\sim 10^{38}$ \ergss, \citealt{hoetal95}),  with an open-ended lower limit \citep{stasinskaetal08}. 
Keeping with the AGN hypothesis for low-luminosity (LL) ELGs  ($L$(\ha) $\lesssim 10^{40}$ \ergss) that fall outside of the \hii\ + TO region, these low-luminosity sources may be associated with massive black holes that are accreting at exceedingly low rates ($\dot{m} \sim 10^{-3} - 10^{-5}$; \citealt{ho05}), well below the rate of luminous Seyfert 1 galaxies ($\dot{m} \gtrsim 10^{-2}$).  
Such sources  represent  a sizable fraction of all galaxies with detected emission lines, whereas Seyfert 1 and 2 represent 1\%\ and 3\%\ respectively of the low-$z$ SDSS galaxies \citep{haoetal05}. If they  are really AGNs, the classical   fueling problem has manifold solutions because of the low accretion rates involved. Accretion material may be due even to mass loss by evolved stars  in the nuclear regions of a galaxy \citep{padovanimatteucci93,heckmanbest14}. 


This interpretation of   LINERs (that are usually of low-luminosity and a large fraction of LL ELGs) as AGNs assumes that line emission is due to gas photoionized by a continuum harder than the one of hot, massive stars. 
The excess X-ray emission of AGNs  leads to an excess of low ionization lines (in the DD case, \oi\ and \sii) with respect to the Balmer lines and high ionization species (in the DD case, \oiiiopt) \citep{veilleuxosterbrock87,kewleyetal01}.  
LINERs, originally defined  from the conditions \oii / \oiii $\ge$ 1 and \oi/\oiii $\le \frac{1}{3}$\ \citep{heckman80} are certainly an heterogeneous class of sources. Especially in the LL   domain, the AGN interpretation is not the only possible one. Sources classified as  low-luminosity AGNs may be true LL AGNs \citep{cozioletal14}, but also galaxies whose emission line gas is ionized by post-asymptotic branch stars \citep{cidfernandesetal10}, or even shock heated \citep[e.g.,][and references therein]{newmanetal14}. The dominant interpretation of LINERs is the one of LLAGNs  due to the  detection in most of these galaxies of a nuclear compact hard-X source \citep{gonzalezmartinetal06,marquezetal07,andersonetal07}. {  Some LINERs can be Compton-thick, and they may be missed in soft-X ray observations (0.2 -- 2 keV, \citealt{gonzalez-martinetal15}). } However, the optical emission line spectrum can be produced by photoionization \citep{ferlandnetzer83,halpernsteiner83} and equally well  by shock heating, as demonstrated since the early 1980s \citep{continialdrovandi83,viegas-aldrovandigruenwald90}.   Shocks, post-AGB photoionization are not exclusive of the AGN scenario and may help explain why the prevalence of these sources is high in elliptical and early type spirals \citep{kauffmann09}. 

If we now turn to clusters,  surveys like the Sloan Digital Sky Survey (SDSS) and WIde field Nearby  Galaxy Cluster Survey (WINGS, \citealt{fasanoetal06})  opened up the possibility of using large samples. However, observations of individual clusters  cover at best $\sim$100 cluster members, and more frequently a few tens.  These numbers are insufficient to satisfactorily sample the relatively rare AGN phenomenon whose prevalence is a strong function of luminosity and AGN class: LINER-type activity may be detected in one third for early type galaxies  at $L \sim 2 \cdot 10^{39}$  \ergss\ \citep{hoetal97,carrilloetal99}, but may not exceed 1\%\ for luminous Seyfert 1 galaxies in the local Universe \citep{huchraburg92,hoetal95}.  In addition, the ionised gas content of cluster galaxies  is expected to be lower than in the field. This makes the study of emission lines in cluster galaxy spectra even more challenging, requiring very high S/N at intermediate resolution  to ensure that faint emission components are detected. 

{ This paper deals the  problems concerning the AGN fraction  in clusters mentioned above, taking advantage of the WINGS spectral database \citep{fasanoetal06,varelaetal09,cavaetal09,morettietal14}.}
The  spectra of WINGS  are adequate in terms of both S/N and dispersion to sample the faint non-\hii\ population (i.e., Seyfert, LINER and TO) at typical equivalent width $\sim$ 1 \AA, but again the relative rarity of detectable nuclear activity requires that  survey clusters  are stacked together  for a prevalence study of AGN classes (\S \ref{spe}).

The sample of cluster galaxies for the present investigation is described in \S \ref{sample}, along with a summary of the main instrumental properties that characterize the spectroscopic data. Measurements  are discussed in \S \ref{meas}, with   data analysis including error estimates and censored data in \S \ref{erranal}. Results, i.e., emission line measurements are reported in \S \ref{res} and are statistically analyzed  in \S \ref{freq} and later subsections. Their discussion (\S \ref{disc}) is mainly focused on a comparison with previous works involving ELGs statistics in cluster (\S \ref{envclu}) and in different environments (\S \ref{envothers}), and especially on the possibility to interpret  low-luminosity non \hii\ ELGs as  predominantly shock heated sources (\S \ref{shock}). We assume cosmological parameters $H_{0}$ = 70 \kms\ Mpc$^{-1}$, $\Omega_\mathrm{M}=0.3, \Omega_{\Lambda} = 0.7$.




\section{Sample definition}
\label{sample}

\subsection{The WINGS cluster sample}

The survey WINGS \citep{fasanoetal06}\footnote{\url{https://sites.google.com/site/wingsomegawings/home}} is an imaging and spectroscopic  study of the brightest X-ray clusters at redshift  $0.04 < z <0.07$  selected from the ROSAT all sky survey .  The basic properties of the 46 clusters considered in this study are reported in Table \ref{tab:clu}. Table \ref{tab:clu} lists cluster name, redshift, velocity dispersion and associated uncertainty in \kms, logarithm of soft X-ray luminosity (0.5 -- 2.0 keV) as in \citet{fasanoetal06},  and the virial radius $R_{200}$\ in Mpc. The following two  columns report, for each cluster, the numbers of sources observed that are spectroscopically-confirmed cluster members ($N_1$) and those that are non-members ($N_0$) following the criterion of \citet{cavaetal09}. The last column label indicates if a cluster was observed from the Northern or Southern hemisphere, with somewhat different instrumental setup (see \S \ref{spe}).


The low-redshift  sample includes clusters with velocity dispersion in the range $500 $ \kms $\lesssim \sigma \lesssim 1100$ \kms, and $43.5  \lesssim \log L_\mathrm{X}  \lesssim 45$\ [\ergss]. These properties place the WINGS clusters { toward the high end} of the  cluster mass  and X-ray luminosity function in the local Universe \citep{bahcallcen92,biviano93,girardietal98,bohringeretal14,reiprichbohringer02,bahcall79,degrandietal99}. Sample standard deviations  rms$_\sigma$  and rms$_{L_\mathrm{X}}$ are 170 \kms\ and 0.3 dex for velocity dispersion and X-ray luminosity, respectively.    As tested for a preliminary report \citep{marzianietal13c}, no strong correlation emerges between  prevalence of ELGs and cluster  velocity dispersion and virial radius in our sample \citep[cf.][] {hwangetal12}. The WINGS clusters show a moderate dispersion in intrinsic properties that underlies the possibility to  join all clusters and form a stacked sample of cluster galaxies. 


\begin{table}
\scriptsize
\setlength{\tabcolsep}{4pt}
\begin{center}
\caption{Clusters belonging to WINGS considered in this study \label{tab:clu}}
\begin{tabular}{lcccccccc}
\\
\hline\hline  
Name &  $z$ & $\sigma \pm \delta \sigma$ &   $\log L _\mathrm{X}$ &   	 $R_{200}$  &  $N_\mathrm{1}$   &   $N_\mathrm{0}$ & N/S  \\  
 & &  [\kms] &  [\ergss]    & [Mpc]\\
 \hline
A1069	&	0.0653	&	690	$\pm$	68	&	43.980	&	1.65422	&	40	&	72		&	S		\\
A119	&	0.0444	&	862	$\pm$	52	&	44.510	&	2.08724	&	158	&	90		&	S		\\
A151	&	0.0532	&	760	$\pm$	55	&	44.000	&	1.83261	&	92	&	176		&	S		\\
A1631a	&	0.0461	&	640	$\pm$	33	&	43.860	&	1.54845	&	125	&	99		&	S		\\
A1644	&	0.0467	&	1080	$\pm$	54	&	44.550	&	2.61226	&	176	&	90		&	S		\\
A1831	&	0.0634	&	543	$\pm$	58	&	44.280	&	1.30299	&	17	&	49		&	N		\\
A193	&	0.0485	&	759	$\pm$	59	&	44.190	&	1.83428	&	39	&	22		&	N		\\
A1983	&	0.0447	&	527	$\pm$	38	&	43.670	&	1.27589	&	45	&	49		&	N		\\
A1991	&	0.0584	&	599	$\pm$	57	&	44.130	&	1.44081	&	35	&	15		&	N		\\
A2107	&	0.0410	&	592	$\pm$	62	&	44.040	&	1.43575	&	36	&	5		&	N		\\
A2124	&	0.0666	&	801	$\pm$	64	&	44.130	&	1.91914	&	29	&	16		&	N		\\
A2169	&	0.0578	&	509	$\pm$	40	&	43.650	&	1.22468	&	37	&	26		&	N		\\
A2382	&	0.0641	&	888	$\pm$	54	&	43.960	&	2.13014	&	152	&	95		&	S		\\
A2399	&	0.0578	&	712	$\pm$	41	&	44.000	&	1.71311	&	125	&	117		&	S		\\
A2415	&	0.0575	&	696	$\pm$	51	&	44.230	&	1.67485	&	98	&	101		&	S		\\
A2457	&	0.0584	&	580	$\pm$	39	&	44.160	&	1.39511	&	56	&	25		&	N		\\
A2572a	&	0.0390	&	631	$\pm$	10	&	44.010	&	1.53178	&	21	&	5		&	N		\\
A2589	&	0.0419	&	816	$\pm$	88	&	44.270	&	1.97818	&	35	&	12		&	N		\\
A2593	&	0.0417	&	701	$\pm$	60	&	44.060	&	1.69955	&	53	&	33		&	N		\\
A2622	&	0.0610	&	696	$\pm$	55	&	44.030	&	1.67205	&	38	&	33		&	N		\\
A2626	&	0.0548	&	625	$\pm$	62	&	44.290	&	1.50593	&	36	&	34		&	N		\\
A3128	&	0.0600	&	883	$\pm$	41	&	44.330	&	2.12231	&	207	&	90		&	S		\\
A3158	&	0.0593	&	1086	$\pm$	48	&	44.730	&	2.61110	&	177	&	101		&	S		\\
A3266	&	0.0593	&	1368	$\pm$	60	&	44.790	&	3.28911	&	225	&	39		&	S		\\
A3376	&	0.0461	&	779	$\pm$	49	&	44.390	&	1.88475	&	92	&	52		&	S		\\
A3395	&	0.0500	&	790	$\pm$	42	&	44.450	&	1.90784	&	125	&	65		&	S		\\
A3490	&	0.0688	&	694	$\pm$	52	&	44.240	&	1.66101	&	83	&	135		&	S		\\
A3497	&	0.0680	&	726	$\pm$	47	&	44.160	&	1.73827	&	82	&	83		&	S		\\
A3556	&	0.0479	&	558	$\pm$	37	&	43.970	&	1.34890	&	114	&	61		&	S		\\
A3560	&	0.0489	&	710	$\pm$	41	&	44.120	&	1.71554	&	117	&	73		&	S		\\
A376	&	0.0476	&	852	$\pm$	49	&	44.140	&	2.05991	&	66	&	22		&	N		\\
A3809	&	0.0627	&	563	$\pm$	40	&	44.350	&	1.35144	&	104	&	91		&	S		\\
A500	&	0.0678	&	658	$\pm$	48	&	44.150	&	1.57561	&	89	&	51		&	S		\\
A671	&	0.0507	&	906	$\pm$	58	&	43.950	&	2.18725	&	20	&	15		&	N		\\
A754	&	0.0547	&	1000	$\pm$	48	&	44.900	&	2.40961	&	126	&	25		&	S		\\
A957x	&	0.0451	&	710	$\pm$	53	&	43.890	&	1.71862	&	65	&	62		&	S		\\
A970	&	0.0591	&	764	$\pm$	47	&	44.180	&	1.83708	&	116	&	67		&	S		\\
IIZW108	&	0.0483	&	513	$\pm$	75	&	44.340	&	1.23989	&	27	&	4		&	S		\\
MKW3s	&	0.0444	&	539	$\pm$	37	&	44.430	&	1.30513	&	32	&	34		&	N		\\
RX0058	&	0.0484	&	637	$\pm$	97	&	43.640	&	1.53951	&	20	&	8		&	N		\\
RX1022	&	0.0548	&	577	$\pm$	49	&	43.540	&	1.39028	&	25	&	19		&	N		\\
RX1740	&	0.0441	&	582	$\pm$	65	&	43.700	&	1.40944	&	20	&	12		&	N		\\
Z2844	&	0.0503	&	536	$\pm$	53	&	43.760	&	1.29425	&	33	&	21		&	N		\\
Z8338	&	0.0494	&	712	$\pm$	60	&	43.900	&	1.71996	&	53	&	33		&	N		\\
Z8852	&	0.0408	&	765	$\pm$	63	&	43.970	&	1.85550	&	53	&	18		&	N		\\
\hline		 		 	 				
\end{tabular}
\end{center}
\end{table}  

\subsection{The galaxy sample of cluster members from WINGS -- SPE (w1)}
\label{spe}

The spectroscopic data are fully described by \citet{cavaetal09}. Here we just recall the main features of the instrumental configurations used for  WINGS spectroscopic survey (hereafter WINGS -- SPE).  

\begin{itemize}
\item Northern clusters: the 4.2 m William Herschel Telescope (WHT) was equipped with the AF2/WYFFOS multi-fiber spectrograph that yielded a spectral resolution customarily of $\approx$6 \AA\ or $\approx$3.2 \AA\ (for one observing run in October 2004) FWHM and 1.6 arcsec fibre diameter. Spectral coverage  ranges from \oii\ to \oi\ for 1305 spectra of this sample sample; 
\item  Southern clusters: the 3.9 m Anglo Australian Telescope (AAT) was equipped the 2dF multifiber spectrograph that yielded a spectral resolution of 9 \AA\ FWHM and a 2 arcsec fibre diameter. Coverage  extends from \oii\ to \sii. 
\end{itemize}

 Spectrophotometric data  are available for 5859 sources in the fields of 46  of the original 77 clusters covered by the survey. Of these, 3514 spectra were of cluster members. All cluster member spectra were then joined to form a stacked sample with 3514 spectra.\footnote{The exact number of sources considered in the following analysis varies depending on the availability of morphological and photometric parameters in the WINGS database. }  The remaining 2345 were field galaxies and allowed the extraction of suitable control samples matching cluster members in luminosity and morphological type (\S \ref{control}).    


Morphological type, luminosity, colors and structural parameters for most of the galaxies of the present sample were  retrieved from the public WINGS database \citep{morettietal14,fasanoetal12,donofrioetal14}. 

\subsection{Non-cluster galaxies (w0) and control samples}
\label{control}

Spectroscopic targets were selected from the available { WINGS} optical B and V photometry \citep{varelaetal09}.  The spectroscopic survey criteria were defined   to maximize the chances of observing galaxies at the cluster redshift without biasing the cluster sample (for a detailed description see \citealt{cavaetal09} and \citealt{fritzetal14}).  The ratio between the sources spectroscopically confirmed to belong to a cluster  $N_1$ with respect to the spectroscopic observations for that cluster i.e., $N_1 / (N_1+ N_0)$\ (where $N_{0}$ is the number of sources with redshift not consistent with cluster membership following \citealt{cavaetal09}) is significantly less than one, as evident from the   columns of Table \ref{tab:clu} that list $N_{0}$\ and $N_{1}$. This means that a sizable sample of non cluster galaxies can be defined. This sample (hereafter denoted also as the ``w0'' sample, or with the subscript 0) -- has been observed with exactly the same instrumental setup used for targets that turned out to be cluster members (the ``w1'' sample, subscript 1),  making it  well-suited for the definition of a control sample (CS). The w1 sample is a ``stacked'' sample that include all 46 cluster spectra. Approximately 170 w0 galaxies     most likely belong to background clusters 
They were identified from the detection of a second red sequence in the color magnitude diagrams of \citet{valentinuzzietal11} displaced from the one of the WINGS cluster, and removed from the w0 sample.

\begin{figure}
\includegraphics[scale=0.35]{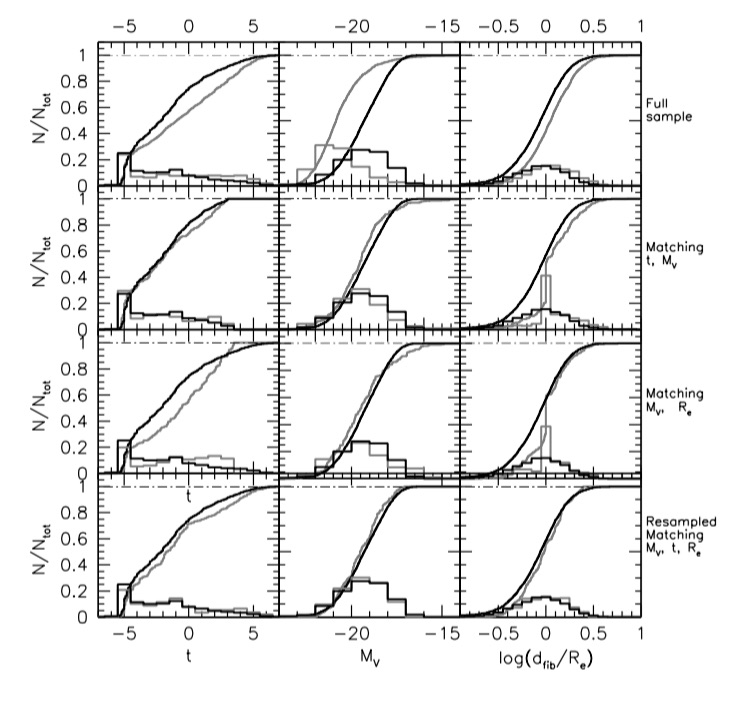}
\caption{Cumulative and differential distribution of  de Vaucouleurs' $t$\ (left panel) , of \mvabs\  (middle), and ratio aperture to effective radius  $\log (d_\mathrm{fib}/$\re), for cluster members (black) and non-cluster galaxies (grey). Top row of panels:  all cluster and non-cluster galaxies included in the present study (i.e., w1 and w0 samples); second row of panels from top : cluster member versus CS   with matched distributions of \mvabs\ and de Vaucouleurs' $t$\ (\cst); third row  from top: cluster member versus CS   with matched distributions of \re\ and de Vaucouleurs' $t$\ (w0$_\mathrm{t,R_\mathrm{e}}$). Bottom row: ``resampled'' CS   matching  \mvabs, $t$,  and \re (\csres).  \label{fig:distr}}
\end{figure}

The redshift distribution 
indicates that the w0 galaxies are mainly background galaxies {  in the redshift range $0.01 \lesssim z \lesssim 0.3$, with distribution mode at $z \approx$ 0.15}, while the cluster galaxies are distributed over a narrow range of redshift {  ($0.04 \lesssim z \lesssim 0.08$)}.   Since  galaxy  emission line properties strongly depend on absolute magnitude \mvabs\  and morphological type, the stacked sample of non cluster galaxies  w0 cannot be compared to the  stacked cluster sample  if systematic differences in ELGs prevalence, emission line luminosity, etc. are under scrutiny.  The distributions of absolute magnitude,  morphological type (de Vaucouleurs' $t$), and ratio between the fibre diameter $d_\mathrm{fib}$\ and \re\ measured in arcsec for non-cluster galaxies   are different from the ones of cluster members (Fig. \ref{fig:distr}), even if evolutionary effects should be relatively minor. The Madau plot \citep{madauetal98,madaudickinson14} shows that evolutionary effects are expected to increase at most the star formation rate by $\delta \log$ SFR $\approx 0.25$\ in the range $0.05 \le z \le 0.3$.  


Non-cluster galaxies still provide a pooling sample   from where control samples matching the luminosity and morphological type distribution of the stacked clusters samples are extracted.    
We constructed a large number  ($\sim 10^{3}$)\ of CSs extracted semi-randomly (as described below) from the stacked sample of non member galaxies to overcome  systematic differences in \mvabs, \re, and $t$.   Non cluster members were randomly selected in \mvabs\ intervals with a  distribution that mimics  the  differential \mvabs\ distribution for cluster galaxies. We then imposed the condition that the \mvabs\  and either \re\ or $t$\ distributions be not statistically distinguishable  from the cluster member distribution at a confidence level larger than 3$\sigma$\ by computing the $D$\ estimator of the Kolmogorov-Smirnov test. The  de Vaucouleurs' morphological parameter  $t$\ is treated as a continuous variable in this context. 

Fig. \ref{fig:distr} shows the cumulative and differential distributions  of $t$, \mvabs, $\log d_\mathrm{fib}$/\re\  with $d_\mathrm{fib} = 2$'' and 1.6'' (for Southern and Northern sample respectively)\ for the full sample and cluster and non member galaxies (top panels), for the full sample of cluster galaxies and one realization of the CSs matching $t$ and \mvabs\  (w0$_\mathrm{t,M}$)  distributions (second row from top), and one matching  $\log (d_\mathrm{fib}/$\re) and \mvabs\ (w0$_\mathrm{t,R}$, third row). 


Extraction of control samples following the procedure described above has several advantages and  drawbacks. 

\begin{itemize}
\item The cluster sample is preserved almost in full. There are limitations introduced to allow for adequate sampling of the w0 parameter space: {  for \cst, 	\mvabs $\ge$ -23.5, $t \le 3.$,  $\log (d_\mathrm{fib}/$\re) $\le 1.5$; for \csres, -17$\ge$	\mvabs $\ge$ -23.5, t unconstrained,  $\log (d_\mathrm{fib}/$\re) $\le 1.4$.} The number of sources is $\approx 3000$ in both cases. 
\item Biases and selection effects are not corrected for, but control samples  are meant to reproduce the same biases and selection effects operating on cluster galaxies. The CSs are however not meant to be representative of a field galaxy population. 
\item The $3\sigma$ conditions is fairly restrictive. Samples are large  so that a $3\sigma$ confidence level is reached  when differences between samples do not introduce a significant bias as shown in Fig. \ref{fig:distr}.  
\item The semi-random extraction procedure  makes possible the realisation of a large number of control samples that are used for  bootstrap  estimates of significance (Appendix \ref{erranal}). 
\end{itemize} 


On the other hand, it proved difficult to extract very large control samples since we could extract at most samples of $\approx$ 300 sources.   In the following we will consider mainly the control samples defined by matching  $t$\ and \mvabs. In this case the median \re\ for cluster and nonmembers samples differ by $\delta \log (d_\mathrm{fib}/$\re)$\approx$0.2, implying a factor of $\approx$ 2.5 in galaxy surface sampled. An ideal control sample would have conditions of statistical indistinguishability concurrently satisfied for $t$, \mvabs,  $\log$\re. Therefore, an additional control sample  has been defined  on the basis of a completeness approach. We considered a 3D parameter space with axes \mvabs, \re, and $t$,   binned in stpdf  $\Delta$ for each variable.  The parameter space volume bins  $\Delta V= \Delta{M_\mathrm{V} \Delta t \Delta R_\mathrm{e}}$  were chosen so that at least two sources were present for  the non-member. We then considered  weighting  factors for the w0 sample sources  given by:  
\begin{equation}
\gimel(M_\mathrm{V},t,R_\mathrm{e}) = \frac{N_{1}(M_\mathrm{V},t,R_\mathrm{e})}{N_{0}(M_\mathrm{V},t,R_\mathrm{e})}, ~~ \forall \Delta V \neq \emptyset 
\end{equation}
Computing this correction factor is relatively straightforward. It has the significant drawback that $\gimel(M_\mathrm{V},t,R_\mathrm{e})$\ becomes $\gg1$ if there are few w0 sources and many cluster ones within a given $\Delta V$.  
The range in the parameter space has no limit on $t$ and \re, and an upper limit in \mvabs $\le$-18 that proved necessary  because fainter sources are almost completely absent in   w0.  Results of one realization of the ``resampled'' CSs (hereafter indicated as \csres)\  are shown in the bottom panel of Fig. \ref{fig:distr}.  


\section{Measurements and Data Analysis}
\label{meas}

\subsection{Spectral modeling}

Faint emission lines ($W \lesssim 1$ \AA) are superposed to the stellar population spectra in a large number of WINGS spectra. They are unresolved  and often fully buried within the absorption associated with the quiescent stellar population.  Sources with strong emission lines  ($W \gtrsim 3$ \AA) are relatively rare (\S \ref{freq}). In a minority of cases the continuum raises toward the blue, emission lines are strong and consistent with high excitation \hii\ emission or with type-2 nuclear activity. Only in two cases we found a type 1 spectrum (\S \ref{lines}). 

Spectral models with synthetic  stellar populations were computed by \citet{fritzetal11} for a large part of WINGS -- SPE galaxies.   Sources without spectral modelling from \citet{fritzetal11} were analyzed using   \citet{bruzualcharlot03} population synthesis templates. The best fit  was achieved by $\chi^2$\ minimisation techniques using the  {\sc IRAF} task {\sc specfit} \citep{kriss94} that allowed each fit to be  carried out interactively.


\subsection{Line intensity, line ratios and equivalent widths}
\label{lines}

\begin{figure}
\includegraphics[scale=0.275]{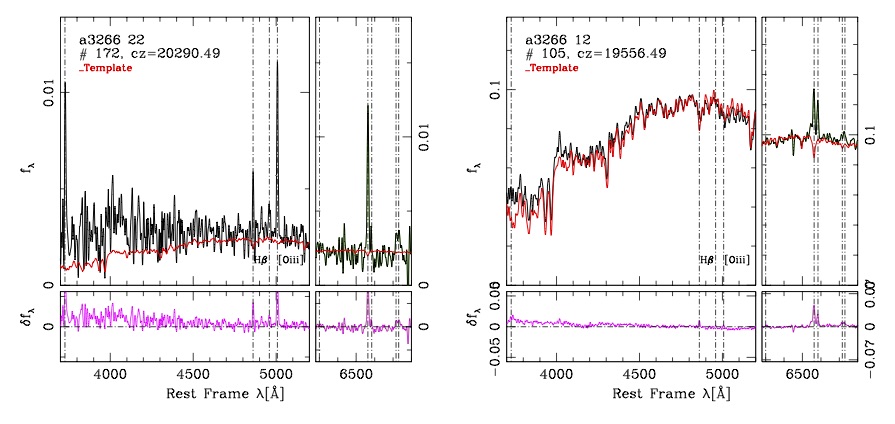}\\
\includegraphics[scale=0.275]{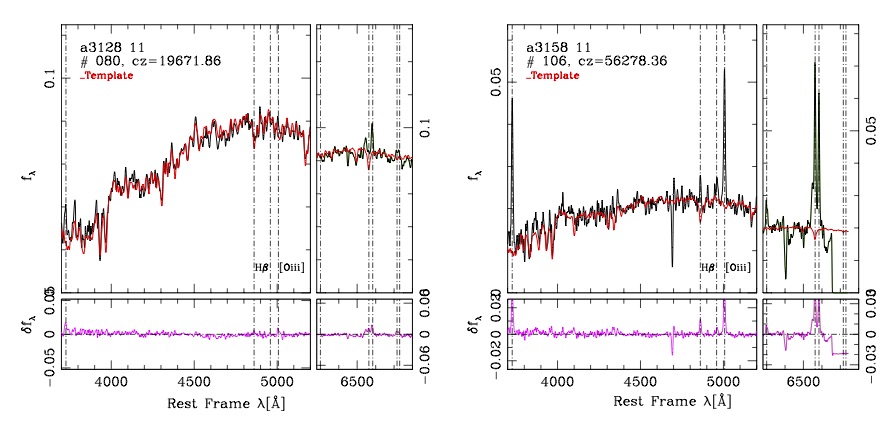}
\caption{Examples of ELG spectra. Top left: the high-ionization \hii\ galaxy  \object{WINGS J043319.43-614141.8}; the TO \object{WINGS\-J043021.56-612848.7}; bottom left: the LINER \object{WINGS\-J032933.77-522659.6}; bottom right the Seyfert 2 galaxy  \object{WINGS\-J034144.52-534221.1}. Red lines trace the adopted stellar template. All of these sources could be entered in the DD with detected diagnostic ratios, and have probability of proper classification $P _\mathrm{i,i}\gtrsim 0.8$. \label{fig:examples}}
\end{figure}

The  typical instrumental properties of   WINGS -- SPE    and the characteristics of the observed spectra justify an approach in which each  spectrum is analyzed as the sum of   stellar emission + faint emission lines whose  width is dominated by the instrumental profile and is not changing significantly from line to line.  The measurement procedure was devised accordingly. The emission components of relevant lines  (\oii, \hb, \oiiiopt, \oi, \ha, \nii, \sii)  were obtained subtracting underlying stellar emission computed by population synthesis.   Examples are shown in Fig. \ref{fig:examples}. Since   population synthesis models often do not satisfactorily  reproduce the whole spectrum from 3700 to 7000 \AA, we applied a two-step procedure to ensure that maximum accuracy is obtained for the \hb\ spectral range. A first  fit was done by {\sc specfit} on a broad range of wavelengths (typically 3800 -- 5500 \AA, with exact values being dependent on  rest frame spectral coverage and presence of zapped regions or contaminated by spikes).  
A second fit was   done restricting the spectral range to the \hb\ region, typically between 4600 and 5100 \AA, to ensure minimisation  of any residual  underlying \hb. This second fit required only minor adjustment in the scaling factor, but it proved to be necessary since the overall best fit did not always yield the minimum $\chi^{2}$\ in the \hb\ spectral region. A small (few tenths of \AA) wavelength adjustment proved also to be  necessary in some cases.  Intensity measures were carried out on the residual spectra i.e., on the original, de-redshifted galaxy spectra minus the scaled population synthesis  model. A second background subtraction was performed on the emission lines in the residual spectra. The intensity measure was taken as the peak value of the line within a fixed window in the residual spectra. 

 
Problematic aspects involved an inaccurate wavelength scale calibration  shortwards of 4000 \AA, because  of the lack of reference lines from lamp and sky. The \oii\ line often appeared displaced with respect to the window used for the automatic intensity line measure.  In such cases, the line was measured interactively by the {\sc iraf} task {\sc splot}. 

An atlas showing the galaxy spectrum, the population synthesis model, and the residual  was created for all the spectra analyzed.  Four examples are shown in Fig. \ref{fig:examples}. The atlas was visually inspected and cross-checked against the measures to ensure the rejection of cases with spike contamination and bad data, as well as identification of cases  that did not meet the assumption on which this procedure is based: sources with broad lines ({  $\gtrsim$ 1000 \kms, as expected for type-1}) and a  mixture of host and non thermal emission. As mentioned,  such sources proved to be {  very} rare: {  3 cases belonging to clusters, and  4 cases not belonging to clusters (their identification is reported in \S \ref{res})}.

\subsection{Criteria for the identification of quiescent and emission line galaxies }

The following selection criterion was adopted to identify sources with emission lines:

\begin{eqnarray}\label{eq:selcrit}
&\Bigg\{ [OII]~ \mathrm{  AND}~ \Big[[OIII]  ~ OR~ H\beta\Big]\Bigg\}& \nonumber \\ 
&~ OR~& \\
&\Bigg\{\left([OIII]~ OR~ H\beta\right)~ \mathrm{  AND}~   \left([NII]~ OR~ H\alpha\right)\Bigg\}&\nonumber
\end{eqnarray} 
 
where a line was considered detected if $I_\mathrm{p}/\mathrm{rms}$\ greater than 3 with $I_\mathrm{p}$ being the ratio between  peak intensity relative to background  and local noise as measured from the rms scatter in a wavelength  range adjacent to the line in consideration. The detection limit on $I_\mathrm{p}/\mathrm{rms}$ has been set to 4 to create a subset of higher quality measures. The criterion can be applied to almost all spectra of the Northern and Southern samples  but  does not ensure  the ability to place each source with detected emission lines in the canonical  diagnostic diagrams. It has the considerable advantage of being less biased than criteria based on a minimum equivalent width of the Balmer lines. Selecting sources on the basis of a minimum equivalent width in \hb\ ($\ge 3$ \AA\ was used in the preliminary analysis of \citealt{marzianietal13c}) clearly biases the selection of ELGs in favor of star forming systems, and may exclude a  population of low luminosity ELGs (for example, low EW sources with \oiii/\hb $>$ 1 and \nii/\ha$>1$) that may not be associated with \hii\ sources.  The analysis of errors on line intensity and line ratios, as well as of censored data is described in Appendix  \ref{andd}.


\subsection{Line flux, luminosity, and SFR estimates \textcolor{red}{}}
\label{flux}

Aperture Johnson magnitudes  $V'$\ and $B'$\  (at fixed width 2 arc sec)  have been  corrected for Galactic extinction (with $A_\mathrm{V}$\  and $A_\mathrm{B}$\ retrieved from NED) and $k$-corrected following \citet{poggianti97}. A line rest-frame luminosity can   be obtained from the line flux $F$\ derived from the aperture magnitudes as in the following equation:

\begin{eqnarray}
L &= &
4\pi D_\mathrm{L}^{2}  W f_{\lambda}    10^{0.4(V'-V'_\mathrm{best})} \label{eq:l}
\end{eqnarray}


where $D_\mathrm{L}$ is the luminosity distance, $W$\ is the line equivalent width. The continuum flux $f_{\lambda}$\ has been   computed from the average of $f_{\lambda, V} = f_{\lambda_{V,0}} 10^{-0.4 V'}$\ and $f_{\lambda, B} = f_{\lambda_{B,0}} 10^{-0.4 B'}$\  for \hb\ (with $ f_{\lambda_{V,0}}$\ and  $f_{\lambda_{V,0}}$ being the 0-point fluxes for the $B$\ and $V$\ bands respectively), and from $f_{\lambda,V}$\ only for \ha. The last factor is the correction for aperture  ($V'_\mathrm{best}$\ is the total isophotal magnitude of the galaxy c.f. \citealt{hopkinsetal03} for \ha), and assumes that the line flux measured within a fixed angular aperture can be scaled by the ratio of the total galaxy flux and of the galaxy flux within the aperture. Estimates of line fluxes and luminosities are therefore expected to be subject to a large uncertainty. 

The star formation rate (SFR) following \citet{kennicutt98} is:

\begin{equation}
SFR =  \frac{  L({H\alpha})}{\eta_{H\alpha}} ~\mathrm{M}_\odot ~\mathrm{yr}^{-1} =    \frac{  L({H\beta})}{\eta_{H\beta}} ~\mathrm{M}_\odot ~\mathrm{yr}^{-1}
\label{eq:sfr}
\end{equation}

where $\eta_{H\alpha} \approx1.27 \cdot 10^{41}$  \msol\ yr$^{-1}$\  erg$^{-1}$\ s, with $\eta_{H\alpha}$  = 2.85 $\eta_{H\beta}$. No correction for internal extinction was applied. 

\section{Results}
\label{res}
\subsection{The Catalog for the WINGS database}
\label{catal}

Relevant quantities extracted from our measurements and analysis are reported in a Table appended to the paper in machine-readable form. The  descriptions of each individual record are provided in Table \ref{tab:table}. Here entries are briefly discussed to explicit their overall meaning. 

\begin{description}
\item[S/N -- ] 
The noise measurement is the rms in the spectral range 5030 -- 5060 \AA, and the S/N is computed by taking the continuum in correspondence of \oiii\ as signal. 

\item[Ratios peak intensity to noise -- ] The ratio peak intensity over local noise $  I_{\mathrm{p}}/\rm{rms}$ is reported for all measured lines. This parameter is used for establishing ELG detection following Eq. \ref{eq:selcrit}, as well as for distinguishing between detections (flag 0) and  non detections in the individual lines (upper limits, flag -1).  
{  The $  I_{\mathrm{p}}/\rm{rms}$ can be used as surrogate of line intensity for nearby lines where the rms noise can be considered constant. Intensity ratios based one line in the \hb\ and and on in the \ha\ range should be viewed with care, due to the uncertain relative spectrophotometric calibration: high S/N spectra often yield $I_{\mathrm{p}}$(\ha)/$I_{\mathrm{p}}$(\hb) $\approx 2$.  }

\item[Equivalent widths --] Rest frame equivalent widths (after stellar absorption correction) are reported in \AA. Errors for detection have been estimated by propagating error on flux and on continuum measures. If the measured equivalent width is below the minimum detectable value computed as a function of S/N, a censorship flag [EW$\_$HB$\_$CENSOR] is set to -1. {  The data are optimized for the \hb\ -- \oiiiopt\ spectral range. W(\hb) should be usually preferred over W(\ha). Even if the \ha\ spectral range in covered in the  Southern sample spectra, the observations  are not optimized at \ha. In several cases the continuum goes down to 0 leading to a divergent W(\ha) value. This is turn affects \ha\ luminosity and SFR estimates based on \ha. }  

\item[Diagnostic ratios -- ]  We report values, uncertainties and censorship flags for the five diagnostic ratios \oiii /\hb, \nii /\ha, \oii/\hb, \oi/\ha, \sii /\ha. In case of detections (censorship flag 0), errors have been computed following \citet{rolapelat94} on the basis of the $I_\mathrm{p}/$rms\ values of each line. Upper limits (censorship flag -1) and lower limits (censorship flag 1) have no associated  uncertainty (nonavailability is coded as -999; {  the code -888 identifies cases in which the spectral range is covered but the ratio can not be computed because both lines are undetected}).

\item[Classification and probability of classification --] We report classification (\hii, LINER, Seyfert, and TO coded as EH, EL, ES, and ET) for sources with detected emission lines and for sources whose emission line ratios allowed to place a data point  in  \ddnii. The data point can be associated with a  detection in two diagnostic ratios, or with lower and upper limit in one or both diagnostic ratios.   For each source in region $i$\ a probability of correct classification $P_\mathrm{i,i}$ is reported along with the probability that the source may be misclassified ($P_\mathrm{i,j}$, with $i \neq j$), and that the correct classification is one of the remaining classes. A revised classification code  is assigned on the basis of the ELG class if  the highest probability of correct classification  occurs in a region  different from the original classification one i.e., if $P_\mathrm{i,j}  > P_\mathrm{i,i}$. The revised classification code has been assigned only for \ddoii\ and \ddnii, and only the latter has been used in the analysis presented in this paper. 
The classifications are reported for the four DDs. However, in examining individual sources, a spectral type  should be assigned on the basis of \ddnii\ which overrules the other DDs. In the case \ddnii\ is not available because the \ha\ range is not covered (Northern sample), then \ddoii\ can be used with \oi/\hb\ to test the possibility of a misclassified LINER, TO, or Seyfert. The  LINER \object{WINGS\-J032933.77-522659.6} and the Seyfert 2 galaxy  \object{WINGS\-J034144.52-534221.1} of Fig. \ref{fig:examples} are classified as \hii\ in \ddoii\ diagram.  {  Individual source classifications make sense only if both diagnostic ratios are uncensored, or for \hii s  whose ratios \oiii/\hb\ and/or \oii/\hb\ and  \nii/\ha\ are upper limits  (which implies $P_\mathrm{i,i} = 1$). 
Otherwise, classification based on censored line ratios have only statistical value  (with the caveats of Appendix \ref{andd})  and, for individual sources,  should be confirmed by additional observations}. 

\item[Flux and Luminosity -- ]   We report specific flux per unit wavelength  at 5000 \AA\ in units of \ergss\ cm$^{-2}$ \AA$^{-1}$,  \ha\ and \hb\ emission line fluxes, along with censorship flags. Due to the tentative nature of flux estimates (\S \ref{flux}), we do not assign an error to specific continuum and line fluxes. The \ha\ and \hb\ luminosity   are  also reported without error assignment.  For these quantities only a detection or censorship flag is given.  

\item[Notes on individual sources --  ]
{  The notes  identify type-1 sources, and sources which belong to background clusters. A comment is added in some cases of instrumental, calibration, or fitting problems, and in the cases a large redshift correction was needed.  }


\end{description}

\subsection{Prevalence of quiescent and emission-line galaxies }

Table \ref{tab:count} lists the number of sources for the full samples of cluster and non member galaxies without restriction on $S/N$ (top),  $S/N_\mathrm{min} = 3$ (middle) and $S/N_\mathrm{min} = 4$\ (bottom).  The code number -999 is used to indicate that all sources are included i.e., that there is no restriction on  W(\hb). The first column reports the samples (w1: cluster member, w0: non-member, w0$_\mathrm{t,M}$, $\overline{\mathrm{w0}}_\mathrm{t,M,R}$). The second column lists the assumed minimum W(\hb) in emission.  Column 3 provides the total number of sources, column 4 ($N_\mathrm{U}$) the number of sources with non-detection. Numbers of sources with detected emission lines are listed in Column 5 ($N_\mathrm{Em}$). Column 6 provides the number of sources that could be placed in the \ddnii\ ($N_\mathrm{DD}$).  The following columns list the numbers of \hii\ ($N_\mathrm{H}$), TO ($N_\mathrm{T}$), LINER ($N_\mathrm{L}$), and Seyfert  ($N_\mathrm{S}$) sources, before and after classification revision using \ddnii\ (\S \ref{andd}).  {  The counts of Table \ref{tab:count} refer  to  diagnostic ratios  computed from detected emission lines. The counts with censored diagnostic ratios (which can be retrieved from the digital table described in \S\ \ref{catal}) yield consistent results.   }


\begin{table*}[htp!]
\tiny
\begin{center}
\caption{Description of fields in the emission line catalog of the WINGS database\label{tab:table}}
\begin{tabular}{lllllll}\\
\hline\hline
  COL  & Identifier & Type & Units & Description \\
\hline
1 & WID& CHAR & NULL & WINGS identifier \\
2 & CLU & CHAR&NULL & Cluster identification code\\
3 & MEM & INTEGER&NULL & Membershift class: 1 member of cluster, 0: non-member\\
4 & IDS & CHAR & NULL & File name with spectrum aperture number \\
5 & SN & FLOAT & NULL & 1-$\sigma$\ S/N ratio measured in correspondence of the \oiii\ line.\\
6 &  R$\_$OII  & FLOAT & NULL & Ratio \oii/$\mathrm{rms}_\mathrm{[OII]}$\\
7 &  R$\_$HB  & FLOAT & NULL & Ratio \hb/$\mathrm{rms}_\mathrm{H\beta}$\\
8 &  R$\_$OIII   & FLOAT & NULL & Ratio \oiii/$\mathrm{rms}_\mathrm{[OIII]}$\\
9 &  R$\_$OI  & FLOAT & NULL & Ratio \oi/$\mathrm{rms}_\mathrm{[OI]}$\\
10 &R$\_$HA  & FLOAT & NULL & Ratio \hb/$\mathrm{rms}_\mathrm{H\alpha}$\\
11 &  R$\_$NII  & FLOAT & NULL & Ratio \nii/$\mathrm{rms}_\mathrm{[NII]}$\\
12 &  R$\_$SII  & FLOAT & NULL & Ratio \sii/$\mathrm{rms}_\mathrm{[SII]}$\\
13 & DETECT & CHAR & NULL & Detection of emission lines following criterion of Eq. \ref{eq:selcrit}.\\
14 & EW$\_$MIN & FLOAT & \AA &Minimum eq. width  $W$\ detectable at S/N reported in Col. 4, from Eq. \ref{eq:ewmin}  \\
15 & EW$\_$HB  & FLOAT & \AA & Rest frame equivalent width of \hb\\
16 & EW$\_$HB$\_$ERR & FLOAT & \AA & Rest frame \hb\ equivalent width error\\
17 & EW$\_$HB$\_$CENSOR & INTEGER & NULL & Rest frame \hb\ equivalent censorship \\
18 & EW$\_$HA  & FLOAT & \AA & Rest frame equivalent width of \ha\\
19 & EW$\_$HA$\_$ERR & FLOAT & \AA & Rest frame \ha\ equivalent width error\\
20 & EW$\_$HA$\_$CENSOR & INTEGER & NULL & Rest frame \ha\ equivalent censorship \\
21 &  R$\_$OIIHB  & FLOAT & NULL & Decimal logarithm of ratio \oii/\hb \\
22 &  R$\_$OIIHB$\_$ERRM  & FLOAT & NULL & Lower error on $\log$ \oii/\hb\\
23 &  R$\_$OIIHB$\_$ERRP  & FLOAT & NULL & Upper error on $\log$ \oii/\hb\\
24 &  R$\_$OIIHB$\_$CENSOR  & INTEGER & NULL & Censorship on $\log$ \oii/\hb\\
25 &  R$\_$OIIIHB  & FLOAT & NULL & Decimal logarithm of ratio \oiii/\hb \\
26 &  R$\_$OIIIHB$\_$ERR  & FLOAT & NULL & Lower error on $\log$ \oiii/\hb\\
27 &  R$\_$OIIIHB$\_$ERR  & FLOAT & NULL & Upper error on $\log$ \oiii/\hb\\
28 &  R$\_$OIIIHB$\_$CENSOR  & INTEGER & NULL & Censorship on $\log$ \oiii/\hb\\
29 &  R$\_$OIHA  & FLOAT & NULL & Decimal logarithm of ratio \oi/\ha \\
30 &  R$\_$OIHA$\_$ERRM  & FLOAT & NULL & Lower error on $\log$ \oi/\ha\\
31 &  R$\_$OIHA$\_$ERRP  & FLOAT & NULL & Upper error on $\log$ \oi/\ha\\
32 &  R$\_$OIHA$\_$CENSOR  & INTEGER & NULL & Censorship on $\log$ \oi/\ha\\
33 &  R$\_$NIIHA  & FLOAT & NULL & Decimal logarithm of ratio \nii/\ha \\
34 &  R$\_$NIIHA$\_$ERRM  & FLOAT & NULL & Lower error on $\log$ \nii/\ha\\
35 &  R$\_$NIIHA$\_$ERRP  & FLOAT & NULL & Upper error on $\log$ \nii/\ha\\
36 &  R$\_$NIIHA$\_$CENSOR  & INREGER & NULL & Censorship on $\log$ \nii/\ha\\
37 &  R$\_$SIIHA  & FLOAT & NULL & Decimal logarithm of ratio \sii/\ha \\
38 &  R$\_$SIIHA$\_$ERRM  & FLOAT & NULL & Lower error on $\log$ \sii/\ha\\
39 &  R$\_$SIIHA$\_$ERRP  & FLOAT & NULL & Upper error on $\log$ \sii/\ha\\
40 &  R$\_$SIIHA$\_$CENSOR  & INTEGER & NULL & Censorship on $\log$ \sii/\ha\\
41 &  CLASS$\_$OII &  CHAR & NULL & Class from location in \oii\ DD\\
42 &  P$\_$OII$\_$HII &  FLOAT & NULL & Probability of HII classification in \oii\ DD\\
43 &  P$\_$OII$\_$LIN &  FLOAT & NULL & Probability of LINER classification in \oii\ DD\\
44 &  P$\_$OII$\_$SEYF &  FLOAT & NULL & Probability of Seyfert classification in \oii\ DD\\
45 &  CLASS$\_$OII$\_$REV &  CHAR  & NULL & Revised class from location in \oii\ DD and probability\\
46 &  CLASS$\_$OI &  CHAR & NULL & Class from location in \oi\ DD\\
47 &  CLASS$\_$NII &  CHAR & NULL & Class from location in \ddnii\\
48 &  P$\_$NII$\_$HII &  FLOAT & NULL & Probability of HII classification in \ddnii\\
49 &  P$\_$NII$\_$TO &  FLOAT & NULL & Probability of TO classification in \ddnii\\
50 &  P$\_$NII$\_$LIN &  FLOAT & NULL & Probability of LINER classification in \ddnii\\
51 &  P$\_$NII$\_$SEYF &  FLOAT & NULL & Probability of Seyfert classification in \ddnii\\
52 &  CLASS$\_$NII$\_$REV &  CHAR & NULL & Revised class from location in \nii\ DD and probability\\
53 &  CLASS$\_$SII &  CHAR & NULL & Class from location in \sii\ DD\\
54 &  FL  &   FLOAT & \ergss\ cm$^{-2}$ \AA$^{-1}$ & Specific flux at 5000 \AA\ \\
55 &  FLV  &   FLOAT & \ergss\ cm$^{-2}$ \AA$^{-1}$ & Specific flux from $V$-band  \\
56 &  L$\_$HB   &   FLOAT & \ergss\  & Decimal log of \hb\ emission line luminosity \\
57 &  L$\_$HB\_CENSOR  &   INTEGER  & NULL  &   \hb\ emission line luminosity censorship flag\\
58 &  L$\_$HA   &   FLOAT & \ergss\  & Decimal log of \ha\ emission line luminosity in units \\
59 &  L$\_$HA$\_$CENSOR  &   INTEGER & NULL  &  \ha\ emission line luminosity censorship flag \\
60 &  NOTES &  CHAR & NULL &  Comments on individual sources \\
\hline
\end{tabular}
\end{center}
\end{table*}

\begin{table*}
\scriptsize
\begin{center}
\caption{Counts of galaxies assigned to different classes from the \ddnii\ \label{tab:count}}
\begin{tabular}{lrrrrrrrrrrrrrr}
\\
\hline\hline
 Sample &  $W_\mathrm{min}$ & $N$&       $N_\mathrm{U}$ &   $N_\mathrm{E}$ &  $N_\mathrm{DD}$ & $N_\mathrm{DD,d}$ &	$N_\mathrm{HII}$ & $N_\mathrm{TO}$ &       $N_\mathrm{Lin}$& $N_\mathrm{Seyf}$&  $N_\mathrm{HII,rev}$ & $N_\mathrm{TO,rev}$ &	 $N_\mathrm{lin,rev}$    &  $N_\mathrm{seyf,rev}$     \\
\hline
\multicolumn{14}{c}{$(S/N)_\mathrm{min}$: no restriction, detection following Eq \ref{eq:selcrit}, $I_\mathrm{p}$/rms $>3$}\\
\hline
w1	&	-999	&	3514	&	2611	&	903	&	801	&	414	&	323	&	58	&	22	&	11	&	330	&	52	&	18	&	14	\\
w0	&	-999	&	2345	&	993	&	1352	&	863	&	448	&	342	&	88	&	6	&	12	&	366	&	63	&	7	&	12	\\ \hline
\multicolumn{14}{c}{$(S/N)_\mathrm{min}$ = 3, detection following Eq \ref{eq:selcrit}, $I_\mathrm{p}$/rms $>3$}\\
\hline
w1	&	-999	&	3461	&	2567	&	894	&	793	&	413	&	322	&	58	&	22	&	11	&	329	&	52	&	18	&	14	\\
w1	&	3	&	448	&	0	&	448	&	448	&	319	&	304	&	9	&	0	&	6	&	304	&	9	&	0	&	6	\\
w0	&	-999	&	2211	&	926	&	1285	&	818	&	434	&	330	&	84	&	6	&	14	&	352	&	61	&	7	&	14	\\
w0	&	3	&	645	&	0	&	645	&	645	&	384	&	316	&	60	&	0	&	8	&	335	&	41	&	0	&	8	\\
w0$_\mathrm{t,M}$	&	-999	&      371  &     144   &      227   &     170    &          96      &     81     &  13  &      1 & 1   &  83     &     11    &       1     &      1\\ 
w0$_\mathrm{t,M}$	&	3	&	147	&	0	&	147	&	147	&	94 	&	81	&	12	&	0	&	1	&	83	&	10	&	0	&	1	\\
$\overline{\mathrm{w0}}_\mathrm{t,M,R}$	&	-999	&	2471	&	1209	&	1262	&	977	&	554	&	487	&	45	&	1	&	21	&	502	&	29	&	2	&	21	\\
$\overline{\mathrm{w0}}_\mathrm{t,M,R}$	&	3	&	756	&	0	&	756	&	756	&	509	&	478	&	25	&	0	&	6	&	493	&	10	&	0	&	6	\\
\hline
 \multicolumn{14}{c}{$(S/N)_\mathrm{min}$ = 4, detection following Eq \ref{eq:selcrit}, $I_\mathrm{p}$/rms $>4$}\\
\hline
w1	&	-999	&	3461	&	2799	&	662	&	607	&	299	&	253	&	25	&	11	&	10	&	253	&	26	&	11	&	9	\\
w1	&	3	&	421	&	0	&	421	&	421	&	260	&	246	&	8	&	0	&	6	&	246	&	8	&	0	&	6	\\
w0	&	-999	&	2211	&	1120	&	1091	&	701	&	319	&	257	&	50	&	3	&	9	&	269	&	40	&	1	&	9	\\
w0	&	3	&	606	&	0	&	606	&	606	&	295	&	250	&	38	&	0	&	7	&	261	&	27	&	0	&	7	\\
w0$_\mathrm{t,M}$	&	-999	&	351	&	163	&	188	&	130	&	67	&	53	&	12	&	0	&	2	&	55	&	10	&	0	&	2	\\
w0$_\mathrm{t,M}$	&	3	&	116	&	0	&	116	&	116	&	62	&	53	&	7	&	0	&	2	&	55	&	5	&	0	&	2	\\
$\overline{\mathrm{w0}}_\mathrm{t,M,R}$	&	-999	&	2471	&	1395	&	1076	&	838	&	371	&	330	&	33	&	0	&	8	&	341	&	22	&	0	&	8	\\
$\overline{\mathrm{w0}}_\mathrm{t,M,R}$	&	3	&	732	&	0	&	732	&	732	&	345	&	323	&	19	&	0	&	3	&	334	&	8	&	0	&	3	\\
\hline
\end{tabular}
\end{center}
\end{table*}

\subsection{A deficit of ELGs in the clusters of WINGS  -- SPE}
\label{freq}

\begin{figure}
\includegraphics[scale=0.275]{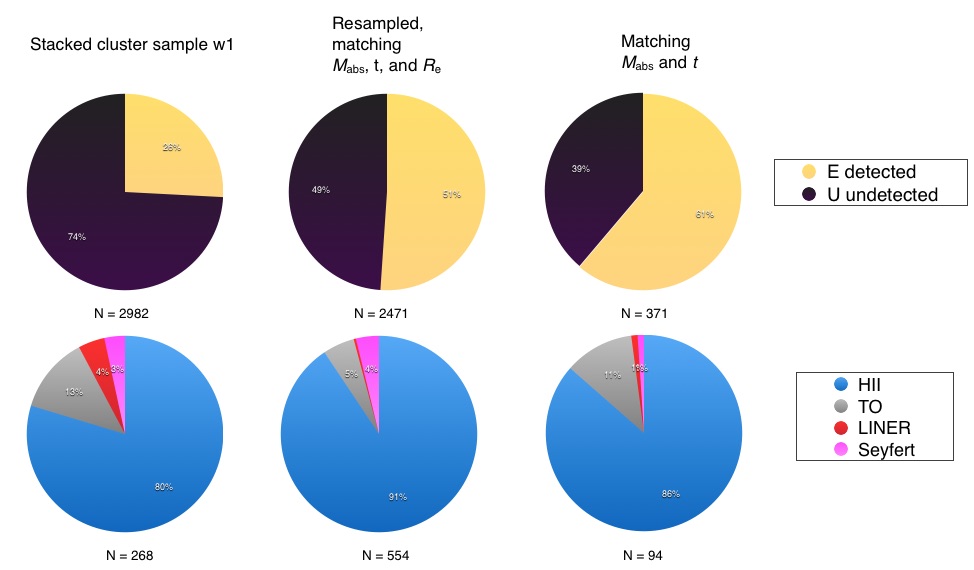}
\caption{Top: fractions of quiescent galaxies (black), and ELGs (yellow), for w1 cluster galaxies (left), one realization of  \csres\ and \cst\ (middle and right). Bottom: prevalence of different ELG classes identified in the \ddnii\ from uncensored diagnostic ratios:   \hii\ (blue), transition objects (pale grey), LINERs (red) and Seyferts (magenta), ordered as in the top row.   \label{fig:pie}}
\end{figure}

The prevalence of ELGs {  is measured by the ratio $f_\mathrm{Em} =  N_\mathrm{Em}/N =  \ N_\mathrm{Em}/(N_\mathrm{U} + N_\mathrm{Em}$),} or by    $R_\mathrm{Em} $= $N_\mathrm{Em}/N_\mathrm{U}$\ = ($N_\mathrm{H}$+ $N_\mathrm{N}$+ $N_\mathrm{E}$)/$N_\mathrm{U}$.  Here  $N_\mathrm{E}$ \  is the number of ELGs with detected emission lines but insufficient data for DD classification, and  $N_\mathrm{N}$ is the number of ELGs that do not show an \hii\ spectrum i.e., $N_\mathrm{N}$ = $N_\mathrm{L}$+$N_\mathrm{S}$+$N_\mathrm{TO} = N_\mathrm{TO}$+$N_\mathrm{A}$, where $N_\mathrm{A}$\ is the number of AGNs (= $N_\mathrm{L}$+$N_\mathrm{S}$). The ratios\ $f_\mathrm{Em}$\ and  $R_\mathrm{Em} $\  are different among cluster members and non-cluster members (Table \ref{tab:count}). The pie diagrams  of Fig. \ref{fig:pie} graphically show a clear deficit of ELGs in clusters. This result holds with respect to the  w0$_\mathrm{t,M}$, w0$_\mathrm{t,R}$, and $\overline{w0}_\mathrm{t,M,R}$, as well as with respect to the w0 sample,  confirming the  preliminary results from a subsample of 1305 WINGS -- SPE sources \citep{marzianietal13c}. 

AGNs (LINERs + Seyferts)   appear to be rare in both cluster and non cluster samples, with prevalence $f_\mathrm{A} \approx 3$\%\ among ELGs. The TO population is sizable in   w1 and control samples   
(Table \ref{tab:count}).  The ratio $R_\mathrm{T} = N_\mathrm{TO}$/$N_\mathrm{H}$\  for revised classes is $\approx  16$\%\ in w1 and   $\approx $ 5\%\ in \cst, and $\approx $ 0.13 \%\  in \csres.  Fig. \ref{fig:pie} and Table \ref{tab:count} indicate that   the ratio $R_\mathrm{N}$ = ($N_\mathrm{TO}$ + $N_\mathrm{A}$)/$N_\mathrm{H}$\   for cluster members  is  larger than or comparable to  CS values.  {   The counts, repeated  applying a slightly different criterion with the conditions  ([OII ~  {  AND}~ ]\hb) | ([OII] ~  {  AND}~ [OIII]) |  [OII]~  {  AND}~ [NII]) | (\hb\ ~  {  AND}~ [OIII]) | (\hb\ ~  {  AND}~ \ha) |  (\ha\ ~  {  AND}~ [NII]), in a logical OR sequence give consistent results.} 

{    Among cluster galaxies we identify  2 Seyfert 1s  (\object{WINGSJ043838.78-220325.0},  \object{WINGSJ060131.87-401646}), and an intermediate type Seyfert (\object{WINGSJ034144.52-534221.1}) One of the 4 w0 Seyfert galaxies  (\object{WINGS J012442.24+085124.4}) was modeled with {\sc specfit} including all relevant components \citep{marzianietal13c}. In the other cases (\object{WINGSJ042931.90-613820.0},
\object{WINGSJ125732.47-173633.1}, \object{WINGSJ132513.37-313137.7}) the broad \hb\ component was not considered in the \hb\ measurements.}

\begin{figure}
\includegraphics[scale=0.425]{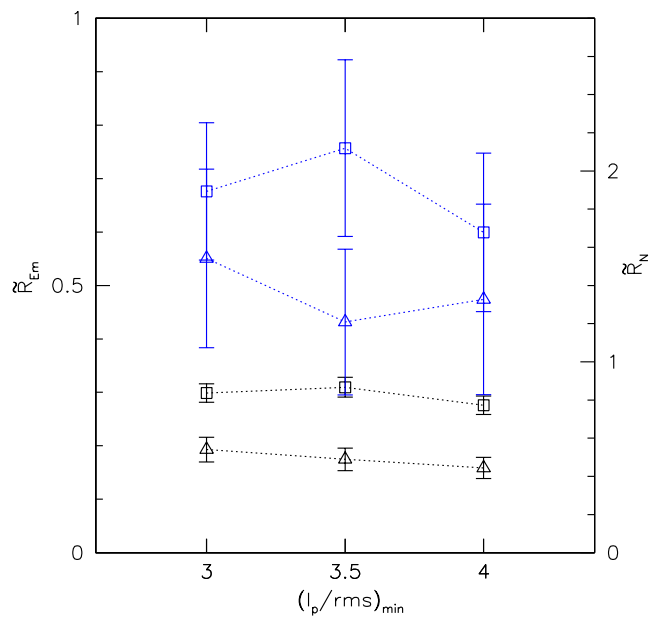}
\caption{Effect of assumed minimum $I_{\mathrm{p}}$/rms on fraction of ELGs, with respect to CSs. The left ordinate shows the ratio  $\tilde{R}_\mathrm{Em}$\ (black lines; squares and triangles indicate normalization by \csres\ and w0$_\mathrm{t,M}$) respectively. The right ordinate shows the ratio $\tilde{R}_\mathrm{N}$\ (blue).
 \label{fig:ratios} 
}
\end{figure}

We tested the result ``robustness'' i.e., that the results are not an artifact of the relatively low minimum $I_\mathrm{p}/$rms (= 3). Fig. \ref{fig:ratios} shows  the ratio between $R_\mathrm{Em}$\ computed for w1 and for w0 and the  CSs (w0$_\mathrm{t,M}$ and \csres): 
$\tilde{R}_\mathrm{Em} = 
\left({N_\mathrm{Em}}/{N_\mathrm{U}}\right)_{1}
/\left({N_\mathrm{Em}}/{N_\mathrm{U}}\right)_{0;\mathrm{cs}}$,
and similarly the ratio $\tilde{R}_\mathrm{N}$, computed dividing by w0 or by one realization of w0$_\mathrm{t,M}$:
$\tilde{R}_\mathrm{N}  =  \left({N_\mathrm{N}}/{N_\mathrm{H}}\right)_{1}/ \left({N_\mathrm{N}}/{N_\mathrm{H}}\right)_{0;\mathrm{cs}}$, 
where the subscript  {0;cs}  indicates the number in either  w0 or one realization of w$0_\mathrm{t,M}$\ and  $\overline{\mathrm w0}_\mathrm{t,M,R}$  for the  minimum $I_\mathrm{p}/\mathrm{rms}$\ equal to 3, 3.5 and 4 (Table \ref{tab:count} and Fig. \ref{fig:ratios};  the case of w0 also confirms the stability of the ratio as a function of $I_\mathrm{p}/\mathrm{rms}$).    The ratio $\tilde{R}_\mathrm{Em}$\ remains $\lesssim 0.3$  for all  $I_\mathrm{p}/\mathrm{rms}$, indicating that the lower frequency of ELGs in clusters is not dependent on the  minimum value of $I_\mathrm{p}/$rms used as  detection criterion.  The ratio\  $\tilde{R}_\mathrm{N}$  is $\approx$ 1.5 if computed for realizations of the CSs and w1.  

\begin{figure*}
\includegraphics[scale=0.4]{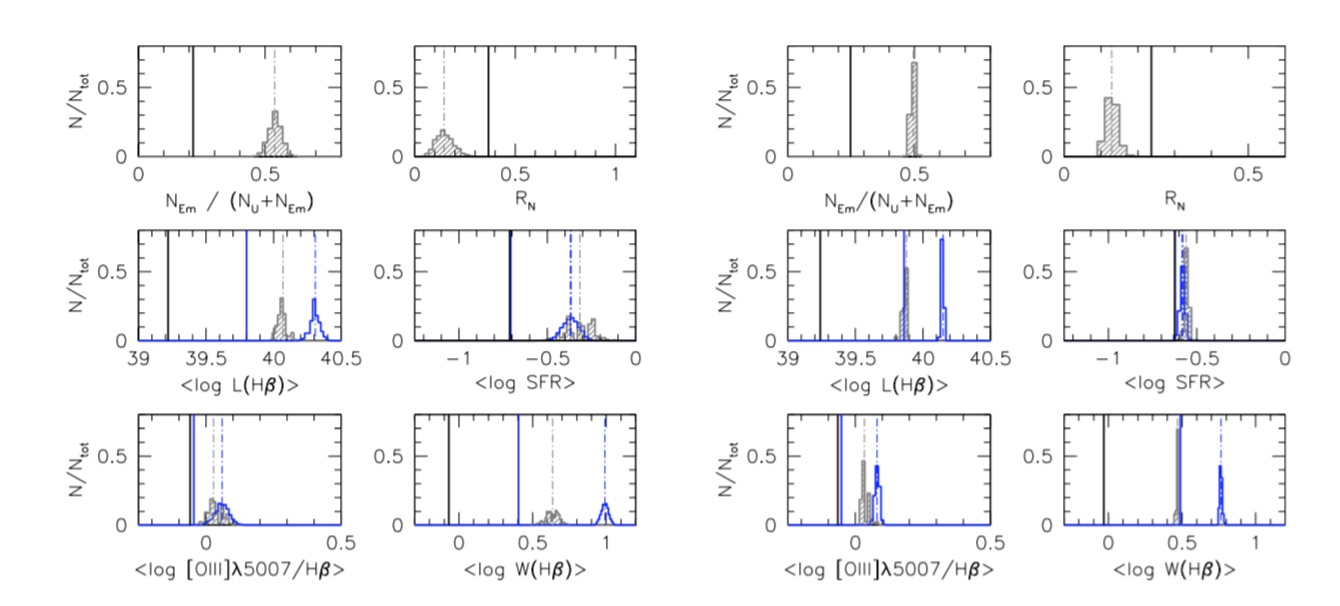} 
\caption{Results of bootstrap simulations. The w1 sample  is compared with 1000  virtual \cst\ CSs of $\approx$ 350 sources   (six leftmost panels), and with 200 virtual \csres\ (six rightmost panels).  The panels show  median values measured  for the cluster  members (thick solid lines) along with the distributions of the bootstrapped sample \cst\ medians  (left) and \csres\ (right) and their medians (dot-dashed lines). Top:  ratio $R_\mathrm{Em}$ (left),   ratio $R_\mathrm{N}$\ (from \ddnii;  right),  for the revised classification (grey). The middle panels show the   luminosity of \hb\ in \ergss\ (left), and the  SFR in \msol\ yr$^{-1}$\ (only for \hii\ sources; right).  Blue histograms and lines refer to average values of detections only, grey histogram and lines show medians and include upper limits. Bottom panel: distributions for  \oiii/\hb\ (left) and W(\hb) (right), with the same meaning of color coding.  \label{fig:boo}}
\end{figure*}

The significance of these results has  been estimated using a bootstrap resampling technique. The upper panel of Fig. \ref{fig:boo} compares the {  median} values for the cluster sample to the distribution of $f_\mathrm{Em}$\   and $R_\mathrm{N}$\  for $\gtrsim$ 1000 \cst\ realizations  (two leftmost top panels), as well as with 200 realization of the resampled   \csres.  
The prevalence  { $f_\mathrm{Em}$\ remains a factor $\approx$ 2 smaller in the cluster sample w1 than in the CSs;   $R_\mathrm{N}$  is {\em larger} by a factor of $\approx 1.5$ (or at least comparable)  in w1 with respect to CSs. In the case of  \csres\  the dispersion of the bootstrapped samples is lower because of the significant number of source repetitions that are necessary with this technique (\S \ref{control}).}

The  lower values of $R_\mathrm{Em}$\    are in agreement with several previous studies (reviewed in \S \ref{disc}), but the larger (or at least comparable) $R_\mathrm{N}$\ for cluster galaxies is an intriguing result. 

\subsection{Prominence of emission lines: weaker, not just fewer} 

Cluster ELGs are not only rarer, they are also weaker.  The number of ELGs is approximately halved if a restriction on $W$(\hb)$\ge$3 \AA\ is introduced. For $W$(\hb)$\ge$3 \AA, the value of  $R_\mathrm{N}$ becomes $\approx 5\%$, down from 0.25 if no restriction on EW is applied. This  immediately suggests  that w1 ELGs  are mostly low luminosity. 
Fig. \ref{fig:boo} compares the medians of $\log L$(\hb) and $\log W$(\hb) for  cluster galaxies and the distributions for the bootstrapped \cst\ (left) and \csres\ (right), including all ELGs regardless of class. Blue lines refer to detections only while grey lines include  upper limits. The $L$(\hb)  differences  are as large as 0.5 dex. $W$(\hb) is systematically lower by $\approx$ 0.5 dex if upper limits are included.  Similar considerations apply to the SFR computed from \hb\ (middle  panels of  Fig. \ref{fig:boo}), for sources classified as \hii\ although the difference in median is a factor $1.5 - 2.5$\ depending on CS. Similar considerations apply to the SFR computed from \ha\ (middle  panels of  Fig. \ref{fig:boo}), for sources classified as \hii.  We also tested the possibility of a systematic difference for the ratio \oiii/\hb. 
The bootstrap analysis again confirms a significant  systematic  difference, with  $\Delta \log$ \oiii/\hb $\approx 0.15 - 0.2$. 
The distributions of $L$(\hb) and $W$(\hb) for clusters and non cluster galaxies (including   all ELG classes)   are found to be statistically different to a confidence level $P \gtrsim 1 - 0.00005$\ by two-sample Wilcoxon generalised tests that include censored data (in this case, upper limits to $L$(\hb) and $W$(\hb)), using the {\sc survival} package implemented within IRAF \citep{feigelsonnelson85}. 

\begin{figure}
\includegraphics[scale=0.62]{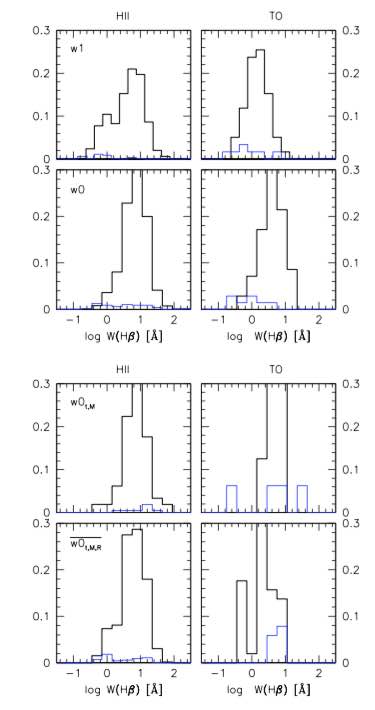}\\
\caption{Distribution of W(\hb) (in \AA) for  sources  classified as \hii\ (left), and TO (right), for w1, w0, one realization of \cst\ and \csres\  (from top to bottom). The  black  histograms trace the distribution of detections,   the thin blue ones  the distributions of upper limits. \label{fig:ew}}
\end{figure}

Fig. \ref{fig:ew} shows the distributions of $W$(\hb) for \hii\ and TOs separately.  The w1 distribution of $W$(\hb) for \hii\ is apparently bimodal or at least strongly skewed.  Clusters galaxies  show  a tail of low-$W$\ \hii\ sources that accounts for the highly significant difference in average equivalent width revealed by the boostrap analysis (Fig. \ref{fig:boo}). TOs, instead, show a single peaked distribution that is clearly shifted toward higher values for w0 and two CSs (apart from fluctuations due to small numbers).  {   A Peto-Prentice generalized Wilcoxon test has been applied  for comparing w1 and  bootstrapped \cst\ and \csres, now separating \hii, TOs, LINERs, and Seyferts, including censored data. The differences in \hb\ equivalent width  are highly significant for TOs and \hii: the Peto-Prentice test statistic is $> 3$\ (TOs) and $>5$\  (\hii) for $\approx$200 bootstrapped samples, implying $P \lesssim 0.025$\ and $\lesssim 0.00005$\ that w1 and control samples  are randomly drawn from the same population. While \hb\ luminosity is significantly higher for \hii\ in the CSs, it is not so for TOs if \cst\ is compared to w1 but the difference become highly significant for \ha\ and for all bootstrap realizations in \csres\ (both \ha\ and \hb). The suggested implication is that  TOs are characterized by a deficit of luminosity per unit mass, rather than a simple general luminosity deficit (as further discussed in \S \ref{mass}). The Peto-Prentice   test indicates lower \oiii/\hb\  (at a $2 \sigma$\ confidence level) for \hii\ and  TOs in more than 90\%\ of the both \cst\ and \csres\ realizations. Differences between w1 and \csres\ in \hb\ EW and luminosity are significant also  for the Seyfert class. If a restriction is done to uncensored line ratios including the \ha\ range, differences in $L$(\hb)\ and W(\hb) become not significant for \hii\ and TOs, alike;  however, this implies  a restriction to the stronger emitters with $W$(\hb) $\gtrsim$ a few \AA.}

\begin{figure*}
\includegraphics[scale=0.90]{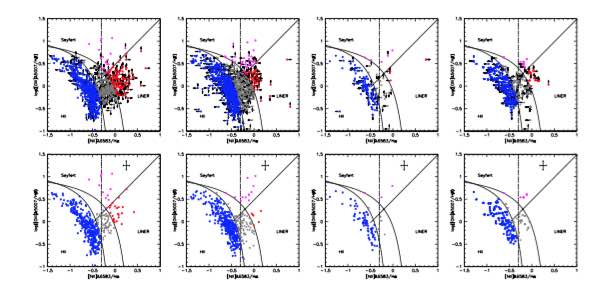}
\caption{\oiii/\hb\ vs \nii/\ha\   diagnostic diagrams. Leftmost panel: DD for galaxies that are  cluster members;  second panel from right: same, for  w0. The third and fourth panels from right show  one realization of \cst\ and of \csres. Blue, red and magenta colors identify \hii, LINERs and Seyferts respectively. Grey data points identify transition objects. Arrows indicate data points for which upper and/or lower limits to the diagnostic emission line ratios are considered. The bottom rows shows panel for detections ordered in the same sequence. {Median errors at a 1$\sigma$ confidence level are shown in the upper right corner of the diagrams with detections only.}   
\label{fig:ddn2}}
\end{figure*}

\begin{figure*}
\includegraphics[scale=0.90]{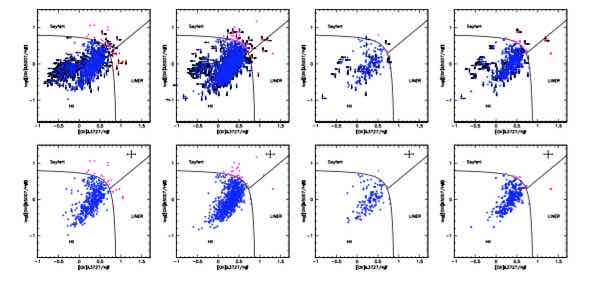}
\caption{Diagnostic diagram  \oiii/\hb\ vs \ vs \oii/\hb. Meaning of panel and symbols and disposition of panels  is the same of Fig. \ref{fig:ddn2}.    \label{fig:ddo2}}
\end{figure*}

\begin{figure*}
\includegraphics[scale=0.90]{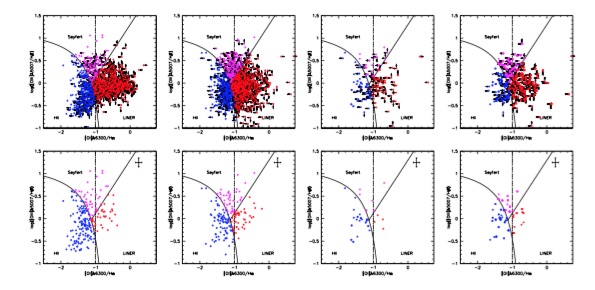}
\caption{\oiii/\hb\ vs \oi/\ha\   diagnostic diagrams. Meaning of panel and symbols and disposition of panels  is the same of Fig. \ref{fig:ddn2}  \label{fig:ddo1}}
\end{figure*}

 \begin{figure*}
\includegraphics[scale=0.90]{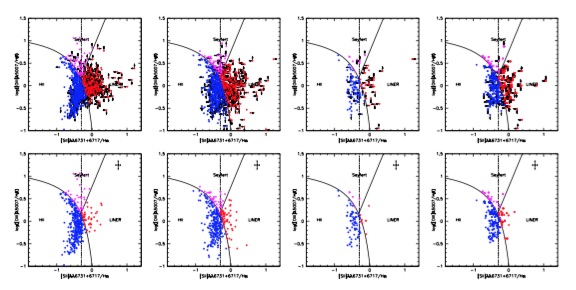}
\caption{\oiii/\hb\ vs \sii/\ha\   diagnostic diagrams. Meaning of panel and symbols and disposition of panels  is the same of Fig. \ref{fig:ddn2}  \label{fig:dds2}}
\end{figure*}

\subsection{Diagnostic diagrams: a  population of low-ionization AGNs  and TOs in cluster}
\label{dd}

The diagnostic diagrams \ddnii, \ddoii, \ddoi, \ddsii\ are shown in Figs. \ref{fig:ddn2}, \ref{fig:ddo2}, \ref{fig:ddo1}, \ref{fig:dds2} respectively.  $S/N_\mathrm{min} = 3$\  is assumed. 
The four panels in each figure show the w1, the w0,  one realization of \cst\ and of \csres.  The diagnostic diagrams that  include censored data  confirm the   trends derived from the diagrams involving detections only. . 

The information provided by each individual diagram is largely independent and subject to different biases (\ddoii: internal reddening; \ddoi\ and \ddsii: very faint lines and predominance of upper limits). For instance, the rarity of LINERs and Seyferts in the \ddoii\ diagram (Fig. \ref{fig:ddo2}) is most likely  due to a bias  i.e., to  a systematic underestimate of  \oii/\hb, since no internal reddening correction was applied. 
Sources in the LINER and Seyfert area of \ddoii\ should be properly classified, but there could be a significant fraction of LINER and Seyferts    that are improperly classified as \hii\ because \oii\ is underestimated.  In  the on-line WINGS database described in \S \ref{catal}  we    report four  source classifications from the different diagrams (record headings are listed in Tab. \ref{tab:table}). 

Revised classifications following the approach of Appendix \ref{andd} are assigned only to sources that enter into the \ddnii\ and \ddoii. The revision affects individual source classification but is not changing  any qualitative statistical result reported  in the paper. {  If no censored line ratios are considered, the count change is more modest, involving only $\sim$ 10\%\ of sources which are mainly reclassified as \hii\ (Tab. \ref{tab:count}). } The number of TOs is significantly reduced after revision {  if censored data are considered} but the relation between w1 and controls is still highly significant because both cluster and control prevalences are reduced by similar amounts.  

Figure \ref{fig:ddn2} emphasizes the abundance of TOs and LINERs in the cluster sample. We stress again that the TO as well as the LINER populations of cluster members almost completely disappear if samples are restricted to large Balmer line EWs ($\gtrsim 3 $ \AA) {as can be deduced from the number counts of Tab. \ref{tab:count}}.   TOs and LINERs are  populations of predominantly weak emitters. 

A large fraction of TOs classified with \ddnii\ is present also in  \ddoi\ and \ddsii\ and classified as LINERs (90 \%), adding further support to the hypothesis that TOs are genuinely different from pure \hii\ regions, at least in the cluster environment. 
LINERs are apparently more frequent than Seyferts in w1, {  if we trust low equivalent width sources close to the detection limit   (\cst\ most frequently does not collect any LINER at all): the average ratio $N_{L}/N_{S} \approx 0.35$ for 1000 realization of \cst, and $\approx$ 1 for w1.   }



\begin{figure*}
\includegraphics[scale=0.90]{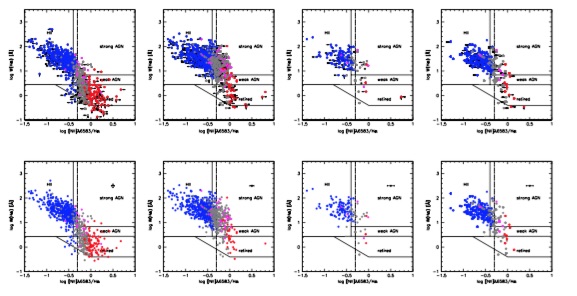}
\caption{Equivalent width of \ha\ W(\ha) vs diagnostic ratio \nii/\ha. Upper panels: sources with detections and upper and lower limits. Bottom panels: sources with detection only.   Leftmost panel  are for sources belonging to clusters,   panels on the right are for control sources: w0  (second from left) and one realization of \cst\ and of \csres.  Dividing lines are drawn according to \citet{cidfernandesetal10}, separating \hii, ``strong AGN,'' ``weak AGN,'' and ``retired'' galaxies. The dot dashed line marks a limiting \nii/\ha\ ratio for extragalactic nuclear \hii\ region.  Color coding identifies \hii, TO, LINERs and Seyfert as in the previous Figures. Black data point identify ELGs with  no entry in  \ddnii.   \label{fig:cid}}
\end{figure*}


\begin{figure*}
\includegraphics[scale=0.90]{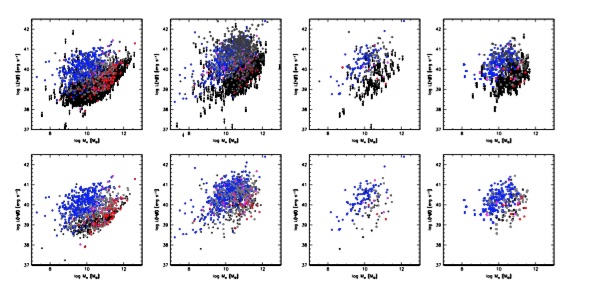}
  \caption{Logarithm of \hb\ luminosity in \ergss\ versus log of stellar mass in solar units, for cluster members  and non-members (middle and right panels), in the following order from left to right: w1, w0, one realization of \cst, one realization of \csres. Upper panels include upper limits and detections, lower panels are for detections only.   Sources of different classes are identified by the same color coding of the previous Figures. 
\label{fig:mlhb}}
\end{figure*}


\subsection{The diagram of \citet{cidfernandesetal10}: an excess of ``retired'' sources}

The emission line weakness of   cluster galaxies is  made even more explicit in the plane $W($\ha) vs \nii/\ha\ (Fig. \ref{fig:cid}; meaning of symbols and arrangement of panels is the same as in the previous figures).  Sources are separated into strong AGNs, weak AGNs, \hii, and ``retired galaxies,'' the latter class including sources for which $ W$(\ha) $\le 2.5$ \AA\ \citep{cidfernandesetal10} i.e., old galaxies with weak emission lines whose spectrum is similar to the one of LINERs \citep{stasinskaetal08,stasinskaetal15}. 


Especially striking is the systematic difference in $W($\ha) for TO between cluster (left) and non cluster galaxies:   $W($\ha) in TOs appears weaker by a factor of 10.  In addition, cluster members  tend to occupy much more frequently the area of retired galaxies which is scarcely populated even in the full sample of non cluster galaxies. The CSs on the right confirm these trends, even if, owing also to the lower $R_\mathrm{N}$\ ratio, relatively few objects populate the area of the diagram with \nii/\ha $\gtrsim -0.3$\ in the nonmember samples. 

A first possibility in the interpretation of TOs is that we are simply witnessing the same scaled-down phenomenon as in non-cluster  galaxies. However,  the diagram $W($\ha) vs \nii/\ha\ argues against the suggestions that TOs in clusters are due to a mixture of gas photoionized by a hot stars and an AGN: in this case, no decrease in line EW is expected. 

\section{Discussion}
\label{disc}

In the following, we  discuss the nature of the non-\hii\ emitters (\S \ref{llelgs} and \S \ref{shock})   without entering the details of individual cluster and  galaxy properties.  We   compare  the present work and other surveys of ELGs for clusters (\S \ref{envclu}),  and works for different surroundings (compact groups and isolated galaxies, \S \ref{envothers}).

\begin{figure}
\includegraphics[scale=0.425]{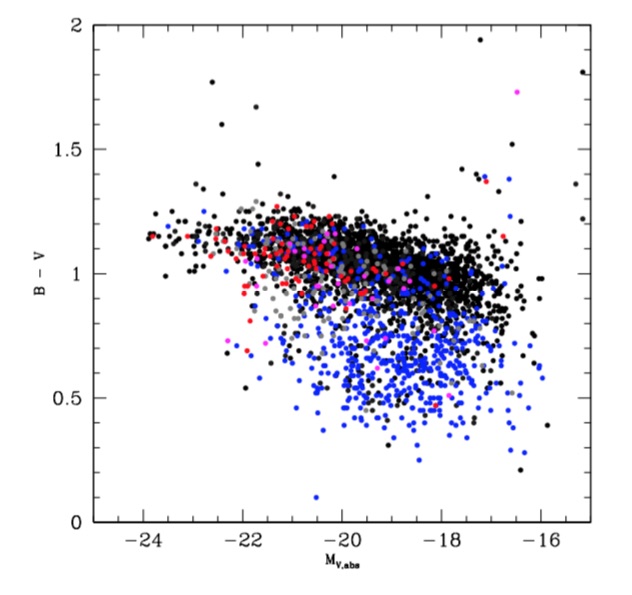}
 \caption{Color magnitude diagram $B - V$\ within a 5 kpc aperture  vs. \mvabs\ for the galaxies of the  cluster sample w1. Color coding is the same as in the previous figure. \label{fig:cm}}
\end{figure}

\subsection{On the nature of non-\hii\ LL ELGs in cluster}
\label{llelgs}

\subsubsection{Almost quiescent line emitters}
\label{mass}

Systematically weaker lines account for the  placement of most ELGs in cluster among ``retired'' galaxies, and for the almost vertical displacement in the diagram of Fig. \ref{fig:cid} \citep[c.f.][]{pimbbletetal13}.  {A change in the amount of ionised gas at a given level of stellar continuum  emission is implied by the definition of EW. We also show $L$(\hb)  vs. stellar mass $M_\star$ in Figs. \ref{fig:mlhb}   (a similar trend hold for \ha, and is not shown), where  $M_\star$\ has been collected from \citet{fritzetal11}, and the mass value is their Mass 1, i.e., the sum of the masses of all stars ever formed and of the remaining gaseous component.}  {An intriguing difference between the w1 and CSs is related to the $L$(\ha)\   distribution of \hii\ sources: a fraction of sources shows comparable values in w1 and CSs but  w1  also shows a significant low-$L$\ tail that is probably associated with gas depletion. This trend is present over the whole $M_\star$\ range considered in Fig.  \ref{fig:mlhb}, and is probably reflected also in  the W(\hb)  distribution of \hii\ soources (Fig. \ref{fig:ew}). } {  The TO $L$(\ha)\ distribution is affected by some very massive hosts at $\log M_{\star}\gtrsim 11$ [\msol] which are poorly represented in the CSs. If the mass range  is restricted to  $9 \lesssim \log M_{\star}\gtrsim 11$, then the $L$(\ha)\ is systematically lower than the CSs, confirming a lower amount of emitting gas per unit mass, as emphasized by the equivalent width trends. A first possibility in the interpretation of TOs is that we are simply witnessing the same scaled-down phenomenon as in non-cluster  galaxies. However,  their interpretation on the basis of a mixed \hii\ + AGN nature is not immediately favored: if this were the case, TOs should show a larger EW than \hii, at least on average.   Due to  their  low equivalent width the wide majority of TOs belong to the $e(c)$\ (moderate to weak emission lines) and $k$ (resembling K stars with no emission lines) in the spectral classification devised by \citet{fritzetal14}. Retired ELGs are not associated with jellyfish galaxies \citep{poggiantietal16}:  jellyfish galaxies belonging to the w1 sample  show prominent emission lines, often with \hii\ spectrum. }

{The color magnitude diagram (Fig. \ref{fig:cm}) $B-V$ vs \mvabs\ for the w1 sample shows that the TOs are mainly located along the quiescent population of early type galaxies (where LINERs are also found) or, to a lower extent in the so-called ``green valley'' where  X and IR selected AGN are usually found \citep[e.g.,][and references therein]{fangetal12}. 

\subsection{Shocks induced by ram stripping?}
\label{shock}

An outstanding  result of this investigation is the placement of most  cluster TO and LINER sources in the area of retired galaxies of Fig. \ref{fig:cid}, along with an $R_\mathrm{N}$\ value that is slightly higher for w0 than for CSs. From  Fig. \ref{fig:cid} we see that this is not the case neither for  sample w0 nor for the CSs. The non-cluster LINER and TO    populations are preferentially placed in the strong and weak AGN region of $\log W$(\ha) $\gtrsim 0.4$\ and  $\lesssim 0.4$, respectively. In the AGN interpretation, this may mean that TOs and LINERs in clusters are simply ``scaled down'' versions of their field counterpart. However, the systematic differences in equivalent width and luminosity and  the relatively unfrequent detection in X rays leave open the possibility that cluster LINER and TOs may be due to  a different physical mechanism. {  Recent works suggest  that X-ray detected AGNs are underrepresented in the cluster central regions \citep[e.g.,][]{koulouridisplionis10,ehlertetal14}, even if cluster LINERs may not recognized as X-ray emitters, as the X-ray detection of individual cluster galaxies is hampered by flux and resolving limits,  and Compton thick LINERs may be undetected in the soft X-ray domain \citep{gonzalez-martinetal15}.}

Observationally, line emission from shock heated  gas with a LINER-like spectrum has been detected in a variety of situations, associated with strongly interacting systems (such as for example  Kar 23) with molecular cloud collisions \citep[e.g.,][]{marzianietal94,marzianietal01,appletonetal06,ogleetal07,monreal-iberoetal10,merluzzietal13}. However, the shock phenomenology is not limited to interacting galaxies.  A warm molecular hydrogen tail due to ram pressure stripping has been detected in  a cluster galaxy, ESO 137-001 \citep{sivanandametal10}. \citet{sivanandametal14} provide imaging in both \ha\ and H$_{2} $\ line  at 17. $\mu$m (H$_{2}$\ 0 -- 0 S(1) transition) for four cluster galaxies, showing a close association between \ha\ emission and part of the H$_{2}$\ emission due to molecular shocks. An especially interesting case in this respect is the one of NGC 4522, where \ha\ and H$_{2}$\ emission extend across the inner galaxy disk.


The mass of ionized gas needed to account for the observed \ha\ luminosity is $M_\mathrm{H^{+}} = L(H\alpha) m_\mathrm{p} / n \alpha_\mathrm{H\alpha}  h \nu_\mathrm{H\alpha}$, where $m_\mathrm{p}$ is the proton mass, $n$\ the number density, $\alpha_\mathrm{H\alpha}$\ the effective recombination coefficient for \ha\ \citep{osterbrockferland06}, $h$\ the Planck constant and  $\nu_\mathrm{H\alpha}$\ the frequency of the \ha\ photons. For the conditions appropriate in the ISM medium  $M_\mathrm{H^{+}}   \approx 2 \cdot 10^{5} L(H\alpha)_{39} n_{10}^{-1}$ \msol\ (with $L(H\alpha)$\ in units of $10^{39}$ \ergss, and $n$\ of 10 cm$^{-3}$),  a modest amount even for the gas-deficient cluster galaxies. Energetically, shock emission from $n$ =  1 cm$^{-3}$  gas would imply a covering factor that can be well $f_\mathrm{c} \sim L(H\alpha) / f_{\Sigma} \pi R_\mathrm{e}^{2} \sim 10^{-1} \ll 1$\ (where $f_{\Sigma}$ is the surface emissivity  $\sim 10^{-1}$ \ergss\  cm$^{-2}$, \citealt{allenetal08}). 

Shock heating models appropriate to \hi\ gas account  for the observed emission line ratios of TOs and LINERs.  Observed emission line ratios, assuming that all line luminosity is due to shocks, can be explained by moderate velocity shocks with precursor (i.e., with ionization in advance of the shock front provided by gas heated in the post-shock zone).  The extension to the right  of data points that enters into the region of LINERs in \ddnii\ (also known as the right ``wing of the seagull'')  can be explained as in Fig. 31 of \citet{allenetal08}, that is as due to shocks with precursor and shock velocity $\lesssim $ 500 \kms.  A grid of models with the same limits in shock velocity   accounts in part   for the distribution of data points in \ddoi, although the observed \oiii/\hb\ ratio is lower than model prediction for the case of shock+ precursor. At face value the \citet{allenetal08} grid computations would suggest shocks without precursor and  a large magnetic parameter that would lower shock compression and hence lead to lower post-shock temperatures. However, we remind that inferences from  \ddoi\ are especially speculative since most measures are upper limits.





Evidences exist that part of the molecular gas content can be stripped by ram pressure, possibly in a sort of progressive ablation at the rims of the molecular gas disk \citep{richetal11,sivanandametal14}.  However, molecular gas content is apparently less affected in cluster environment, or, at least, stripped less efficiently than atomic gas \citep{bosellietal14a}, probably because of the larger extent of the {\sc Hi} disc, and of the different hydrodynamical  effects of the ICM on atomic and molecular gas. The evaporation timescale $t_\mathrm{evap}$ depends  on cloud size, density and ICM temperature: $t_\mathrm{evap} \propto n_\mathrm{c} r_\mathrm{c}^{2} T_\mathrm{ICM}^{-2.5}$\ \citep[][]{cowiemckee77}, and $t_\mathrm{evap} $\   is much longer for molecular clouds ($n_\mathrm{c} \sim 10^{6}$ cm$^{-3}$) than for atomic clouds ($n_\mathrm{c} \sim 10^{1 - 2}$ cm$^{-3}$) of the same size. In the molecular  case, $t_\mathrm{evap} $  can easily exceed the dynamical timescales for cluster galaxies   and may even exceed the Hubble time. 
 Thus the molecular gas could represent a permanent, or at least a long lived-reservoir of gas in cluster galaxies, leading to non negligible star formation and to a mixed Starburst/shock  or to a shock phenomenology in a large fraction of cluster galaxies. Mechanical heating and induced star formation may replenish  cluster galaxies of atomic gas from molecular clouds \citep[e.g.,][]{hidakasofue02}.



\subsection{Comparison with previous works for cluster galaxies}
\label{envclu}

A  quantitative comparison with previous studies is not easy, because  of differences in luminosity, morphology,  and redshift in the galaxy samples,  as well as because of  data heterogeneity  i.e., of differences in emission line detection limits associated with S/N and spectral resolution. We limit ourselves to elementary considerations on $R_\mathrm{Em}$, $R_\mathrm{N}$, type--1 AGNs, and systematic differences in line luminosity. 

\paragraph{Prevalence of ELGs.} The notion that ELGs in cluster are significantly less common than in the field and other environments has been  consolidated by studies spanning more than 30 years   \citep[e.g.,][]{gisler78,balickheckman82,dressleretal85}. This basic result has remained unchallenged until now, and is confirmed by the present works for the X-ray bright clusters of the WINGS -- SPE survey.   There is an overall consistency concerning qualitative trends with the \citet{hwangetal12} results for  clusters regarding  the low prevalence of ELGs in clusters. 


\paragraph{AGN+TO fraction $f_\mathrm{N}$.}    The prevalence of non-\hii\ ELGs,  $f_\mathrm{N}$\ (not normalized by $N_\mathrm{H}$\ i.e., $f_\mathrm{N} = f_\mathrm{AGN}$, where $f_\mathrm{N}$\ is as defined by \citealt{hwangetal12}) in the WINGS cluster sample is only slightly lower than in w0 and CSs.   The data of Table \ref{tab:count} show that, after redistributing unclassified ELGs, $ f_\mathrm{N}\approx$ 5\%\ for w1 is comparable to the prevalences  $  \approx$ 4\%\ and 6\%\  for \cst\ and \csres\ respectively (that become $\approx$ 10\%\  if identifications based on censored ratios are included). The WINGS values including censored data are in  agreement with  recent $ f_\mathrm{N}$ determinations: if the C1 and C2   samples of \citet{hwangetal12} are joined, their $f_\mathrm{AGN}$\ is $\approx  15$\%.  If a restriction to strong emission lines is introduced ($W_\mathrm{min} = 3$ \AA), the non-\hii\ prevalence in cluster is lower than in the CSs by a factor $\approx$ 1.5 (2.5 if censored data are included), in qualitative agreement with older studies that found a  lower prevalence in cluster with respect to less dense environments.   The non-\hii\ sources we detect as TOs  are mainly of low equivalent width, to the point of being located in the retired galaxy region of Fig. \ref{fig:cid}, and their contribution might have been missed in past surveys.  

\paragraph{$R_\mathrm{N}$}  The $R_\mathrm{N}$\ ratio, like $f_\mathrm{N}$,   
can be measured with some precision because it involves a  large number of galaxies.  
 The $R_\mathrm{N}$\ values found by the present work and by \citet{hwangetal12} are different but consistently high ($\approx$ 0.26\ vs. 0.86, again joining samples C1 and C2 of \citealt{hwangetal12}).   The values  probably  reflect  differences in the application of diagnostic diagrams, and in cluster-centric distance coverage. 

\paragraph{``True'' (or ``pure'') AGN.}  The prevalence of AGNs (``true'' AGN excluding TOs) in our sample is not statistically different from the ones in the CSs: $\approx$ 1\%\ for uncensored diagnostic ratios (Tab. \ref{tab:count}), and $\approx$ 3\%\ if censored data are included. This value is  consistent with the prevalence found for the C1+C2 sample of  the recent \citet{hwangetal12} work.  

\paragraph{Type 1 AGN.}  We  found 2  type-1 sources (\object{WINGSJ043838.78-220325.0} and \object{WINGSJ060131.87-401646}), i.e. $\frac{1}{2}$ of {  the detected  type-2 AGN, as expected for our sample size on the basis of orientation-based AGN unification schemes} \citep{antonucci93}.   If the host galaxy is gas-rich, even slight tidal disturbances (i.e., disturbances not as remotely dramatic as mergers like harassment \citep{mooreetal96,hwangetal12} can trigger nuclear activity.   While star formation and type 2 nuclear activity may be concomitant, with type 2 activity delayed with respect to star formation but observed as contemporary in a large fraction of systems (and this may account for at least some TOs), type 1 activity may be triggered by interaction but may be associated with a significant delay \citep[e.g.,][]{krongoldetal03,koulouridisetal13,villarroelkorn14} so that the close environment may not appear different from the one of non-active galaxies. In this case, gaining evidence of past interactions   requires a much more thorough analysis of the type 1 host morphology and environment.  {  On the other hand, active galaxies in clusters are presumed to be gravitationally bound to the cluster, so that, if the evolutionary scheme is correct, they may still belong to the cluster after the delay needed for the onset of type-1 activity. }

\subsection{Comparison with previous works on different environments: compact groups and isolated galaxies}
\label{envothers}

\paragraph{Compact groups.} \citet{martinezetal10}  found a large fraction of ELGs in a sample of Hickson compact groups (HCG): almost two thirds of their sample show  emission lines, and 2/3 of ELGs are AGNs + TOs.  These values can be compared to the completeness-corrected frequencies in  WINGS: 31\%\ of ELGs, with about one third of them AGN + TO.    The WINGS data therefore confirm that the AGN fraction  is  significantly lower in clusters than in galaxy groups, with fractions doubling from clusters to groups. This result apparently holds both if AGNs are X-ray selected  and if  optical DDs are used. The AGN fraction in HCG galaxies with  $L_\mathrm{X, 0.5-8.0 keV}  \ge 10^{41}$ \ergss\ is $0.08^{+0.35}_{-0.01}$,   higher than the $\approx$ 5\% fraction in galaxy clusters \citep[][]{arnoldetal09,tzanavarisetal14}.  The  value of $R_\mathrm{N} \gtrsim 1$\  is higher for compact group   than   for WINGS cluster galaxies\ \citep{martinezetal10,sohnetal13,bitsakisetal15}.   \citet{martinezetal10} suggest that the level of activity in HCGs is characterized by a ``severe deficiency of gas,'' and indeed the L(\ha) luminosity  is comparable to the one found in our cluster sample.  Apparently, compact groups share some of the deficits found for clusters although at a less extreme level.  {  In this context, it is interesting to note  that many compact group galaxies contain molecular gas that is not forming stars efficiently \citep{alataloetal15}, and that   evidence of shocks has been found in the emission line ratios   of the late-type galaxies in compact groups \citep{bitsakisetal16}. }

The absence of type-1 AGNs has been noted  in the dense environments of dynamically old  compact groups \citep{cozioletal00,martinezetal08,bitsakisetal15}.  In these dense environments, accretion processes in the nuclei of galaxies may be  significantly affected by  a hot intergalactic medium revealed through its  X-ray emission. Absence of type 1 sources in dense environment {  has led to the suggestion that  unification schemes may be} dependent on active nucleus luminosity  and/or environment \citep[e.g.,][]{dultzinhacyanetal99,dultzin-hacyanetal03,krongoldetal03}, or  at least some LLAGNs may not be real AGNs.


\begin{figure}
\includegraphics[scale=0.7]{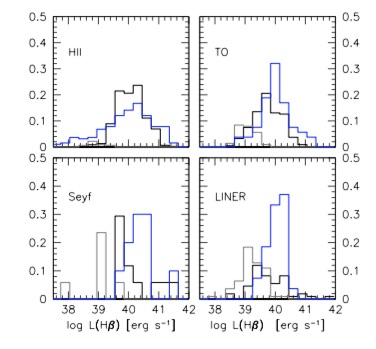}
 \caption{Distribution of L(\ha) for the WINGS sources (w1 sample), for \hii, TOs, Seyferts and LINERs (black line), and for the sources analysed by \citet[][blue]{sabateretal12}. Grey lines trace the distribution of upper limits for the w0 sample.  {  Object classified with censored emission line ratios are included in the histogram.}\label{fig:saba12}}
\end{figure}

\paragraph{Isolated galaxies.}  Isolated galaxies as defined by \citet{verdes-montenegroetal05} represent in many ways the ``opposite'' environment from cluster galaxies, in terms of galaxy surface density:  no companion galaxy of diameter $d$\ within 1/4 and 4 diameter of the primary should lie within 20$d$ from the primary. Isolated galaxies are believed to be sources whose properties are due to internal secular evolution and are linked to formative evolution. External influences are expected to be minimized, since they isolated galaxies may  not have  interacted with a  neighbor of significant mass in the past $\approx$ 3Gyr.  Most recent studies detect emission lines in almost all galaxies \citep{varelaetal04,hernandez-ibarraetal13}. The fraction of AGN depends on morphological type and luminosity, being higher for earlier morphological types and at high luminosity  \citep{sabateretal12,hernandez-ibarraetal13}, and is about 30 \%\ (without TOs) -- 40 \%\ (with TOs).  An even higher fraction ($\approx$ 2/3 of sample) of AGNs + TOs was found from the analysis of an SDSS based sample of $\sim $ 200 galaxies \citep{cozioletal11}.   {In the luminosity domain of WINGS ELGs, $10^{39} - 10^{41}$ \ergss,}  AGNs and TOs are found at a percentage slightly lower than the one of \hii, with $R_\mathrm{N}$\    $\approx $ 0.7   \citep{varelaetal04,sabateretal12,hernandez-ibarraetal13}.  The   L(\ha) distributions for the 4  ELG classes  for the isolated galaxy sample of \citet[][classification and line fluxes are retrieved from the Vizier catalog J/A+A/545/A15 associated with the paper]{sabateretal12} and  for WINGS {   show that   WINGS TOs and LINERs are systematically less luminous} (Fig. \ref{fig:saba12}). 




In summary, comparison of data related to different environments   is not easy because of sample differences in luminosity and morphology. The tentative\ analysis of this paper indicates that there is a sequence of decreasing frequency and luminosity of ELGs from isolated galaxies to group and to cluster environments, holding in the absolute magnitude domain  $-23 \lesssim $ \mvabs\ $\lesssim -18$. The low luminosity of the emission lines makes a mechanism like shock appealing to explain, at least in part,  TOs and retired ELGs in the cluster environment.  {  The 3 type-1 AGNs identified in the WINGS clusters are at clustercentric distances $\lesssim 0.5$ of the virial radius, where a significant effect from ICM is expected. As it is not possible to draw general inferences from few sources, the type-1   detections in the inner regions of the WINGS clusters only suggest  the need for more   focused studies  to analyze how  hot inter-galactic medium may influence  accretion processes.  }
 



\section{Conclusion}
\label{conc}

This paper presents the emission line classification for most galaxies of the WINGS -- SPE survey involving 
 X-ray luminous clusters in the redshift range. Among the immediate results of the analysis are the emission line intensities, equivalent width,   estimates of fluxes, and assignment of a class on the basis of diagnostic diagrams. Sample analysis included  censored data with a rigorous treatment when a two sample comparison was carried out, and with an heuristic approach in the assignment of classification probabilities in the 2D diagnostic diagrams.  The present analysis  relies on  ad hoc control samples that allowed to test systematic differences between cluster and field galaxies in well defined parameters such as $R_\mathrm{Em}$, $R_\mathrm{N}$, line equivalent widths  and luminosities. Field galaxies were used to build control samples with statistically undistinguishable  morphology mix, luminosity and ratio aperture-to-\re\ distribution, and therefore cannot be considered representative of a field galaxy population at low surface density.

The new analysis  adds to the view of galaxies in clusters several results: \begin{enumerate}
\item a confirmation of the long-held notion that ELGs are less frequent in the cluster environment; 
\item detectable line emission is not only rarer, but also weaker, implying a lower amount of ionised gas per unit mass, and  a lower star formation rate if the source is classified as \hii\ region.
\item The presence of a sizable population of sources showing spectra of TOs and LINERs. TOs and LINERs  are more frequent than, or at least as frequent as  in the CSs  with respect to the \hii\ sources,  although they show a much lower W(\ha) than in the CSs,   by a factor    $\gtrsim$2 -- 3. The effect is well-illustrated in Fig. \ref{fig:cid}. 
 \end{enumerate} 
 A number of possible mechanisms can explain the LL TOs and LINERs: true low-luminosity nuclear activity, but also ionization by evolved PAGB   stars, and shocks. Shocks associated with the interaction between the galaxy atomic gas and the ICM provide  emission line ratios in agreement with the observed ones. The phenomenon may be relatively long lived if the molecular gas in the disk of galaxies can act as a reservoir. 

The relation between LL activity and cluster substructure  \citep{ramellaetal07} and other properties of individual clusters (such as the entropy profile)  will be investigated in an eventual paper. Such investigation will benefit of a wider coverage of the outer cluster regions  that is being provided by the ongoing extension of WINGS: OMEGAWINGS \citep{Gullieusziketal2015}.  









\begin{acknowledgements}
P.M. acknowledges the kind hospitality of the   IAA-CSIC at Granada where an early part of this work was done, along with the Junta de Andaluc\'\i a, through grant TIC- 114 and the Excellence Project P08-TIC-3531, and the Spanish Ministry for Science and Innovation through grants AYA2010- 15169 for supporting her sabbatical stay in Granada.
\end{acknowledgements}

\bibliographystyle{apj} 

\begin{appendix}
\section{Error analysis and treatment of censored data}

\label{erranal}


Errors on the logarithm of  line intensity ratios   are in general symmetric  since errors on intensity ratios follow   log-normal distributions. In the present paper, errors   were  estimated   following the prescription of \citet{rolapelat94} that   assumes a proper lognormal distribution for the case of modest values of $I_\mathrm{p}/$rms ($I_\mathrm{p}/$rms $\lesssim 10$, as it is the case of most of our measurements). Errors on  logarithm of  line intensity ratios are significantly asymmetric in the case  the two lines have   different  $I_\mathrm{p}/$rms, and one of them has $I_\mathrm{p}/$rms $ \lesssim 10$. 

\setcounter{equation}{0}
Upper and lower limits were included in the analysis. Emission lines whose $I_\mathrm{p}/$rms is below the limiting values (either 3 or 4) were considered not detected, and an upper limit was set right at the limiting $I_\mathrm{p}/$rms value. Both upper and lower limits in line ratios should be included  in the intensity ratio and diagnostic analysis  in order to fully take advantage of the generality of the selection criteria of Expression \ref{eq:selcrit}  since \hb\ and \ha\ intensity appear  at the denominator. 


Considering only detections would be equivalent to set an ill-defined limit to the \hb\ equivalent width. The broad range of  S/N  present in the WINGS  spectra  
introduced a censoring on emission line detection that is dependent on the line equivalent width. Measuring the equivalent width on mock spectra built with (1) the stellar template of a quiescent population (i.e., the most likely case), (2) an artificial unresolved emission line meant to mimic \hb, and (3) Gaussian noise show that the minimum equivalent width as a function of S/N is given by 

\begin{equation}
W_\mathrm{min} \approx 18.2  \left(\frac{I_\mathrm{p}}{\mathrm{rms}}\right)_{\mathrm{min},3}  \left(\frac{\mathrm S}{\mathrm N}\right)^{-1.10}~~ \mathrm \AA \label{eq:ewmin}
\end{equation}

where the relation is  given for $\left(\frac{I_\mathrm{p}}{\mathrm{rms}}\right)_{\mathrm{min}} = 3$.  The signal to noise ratio  has been measured in correspondence of \oiii, and is meant to be representative of the overall S/N of  each spectrum.   A  S/N $\approx$ 6 would correspond to a minimum $W_\mathrm{min} \approx 3$ \AA. This equivalent width limit  was applied in our preliminary analysis \citep{marzianietal13c}; however from the distribution of  S/N vs $W_\mathrm{min}$ \ we deduce that a large fraction of sources with   detected emission in higher S/N spectra was not considered (Fig.  \ref{fig:snw}).  

\begin{figure}
\includegraphics[scale=0.078]{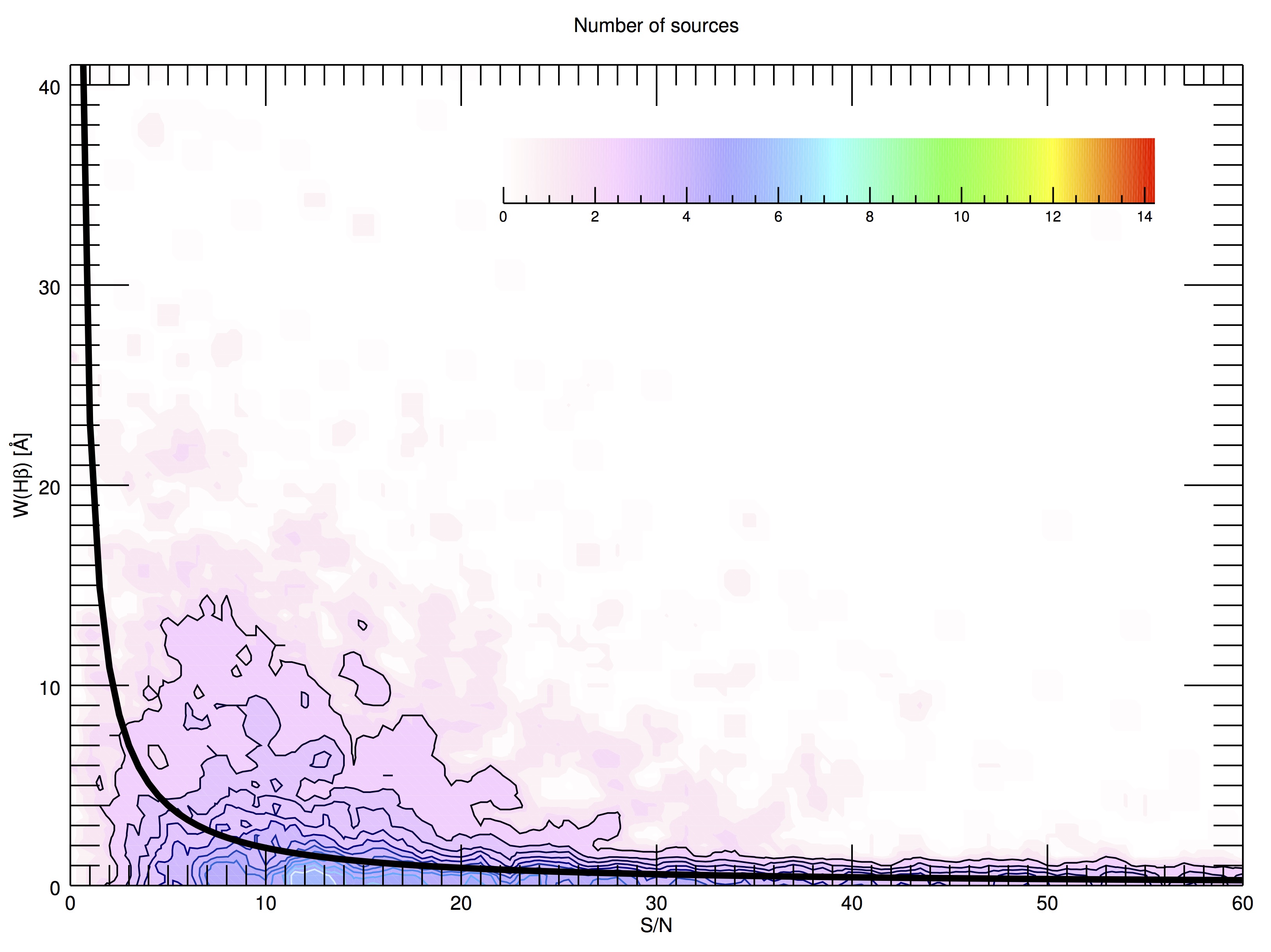}\\
\caption{Minimum W(\hb) (thick line) and the distribution of W(\hb) in the plane W(\hb) vs S/N. Isoplets represent source numbers for the  sample obtained joining w0 and w1,  color-coded as shown by the bar at the top right corner of the plot.\label{fig:snw}}
\end{figure}




\subsection{Analysis of   diagnostic ratios and diagrams}
\label{andd}

The four diagnostic diagrams based on pairs of emission line ratios and  employed in this study  (\oiii /\hb\  vs \nii / \ha, hereafter \ddnii; \oiii / \hb\ vs \oii / \hb, hereafter \ddoii; \oiii / \hb\ vs \oi /\ha, hereafter \ddoi; \oiii / \hb\ vs \sii / \hb, hereafter \ddsii)   are not equivalent. Each of them provides different information on the physical conditions of the emission line gas. In addition, the interpretation  is affected by  instrumental problems specific to each diagram: the \ddoii\ is efficient in detecting ELGs since it involves the strong \oii\ doublet. However, the use of this line for diagnostic analysis necessitates a reliable correction for internal extinction. For the $\approx$1300 sources that lack coverage of \ha\ (Northern sample), this correction cannot be computed. 
The \ddnii\ is by far -- from the point of view of measurement robustness -- the most valuable diagram: it is not strongly affected by internal extinction and involves the strongest lines observed in the spectral range of WINGS -- SPE. It also allows for the finest classification, separating objects in four classes (\hii, transition objects, LINERs and Seyferts, \citealt{kewleyetal01,kewleyetal06}, \S \ref{dd}). 
Low ionization lines like \oi\ and \sii\ are enhanced in an extended post-shock recombination region (analogous to the ``partially-ionised zone''  created by   the AGN soft X-ray radiation, \citealt{krolik99}, and references therein). In a low luminosity context their detection may indicate well either shock or AGN photoionization as a production mechanism. They are especially useful to test the nature of TOs  (\citealt{veronetal97}). However, the \oi\ and \sii\ lines, even if enhanced, remain rather weak, and are often of uncertain detection with WINGS - SPE data. Their use for individual sources should be therefore strictly restricted to detections.

 
 \subsubsection{Assigning the probability of classification}

 A quantitative analysis of the diagnostic diagrams  requires (1) a proper consideration of the source position in the DD  with respect to the limits drawn   to distinguish different ELG classes; (2) an adequate treatment of uncertainties and  upper limits, aimed also at providing a probability that a source is properly classified. 
This approach is  especially needed since: (1) there is a large clustering of sources close to the dividing lines, for example  at the boundary between Seyfert and \hii\ regions. 
(2) TOs are located in a narrow strip between \hii, Seyferts and LINERs. 

We  considered the probability\ $P_\mathrm{i,j}$\ that a source in region $i$\ could be classified   as belonging to   region $j$\ in a diagnostic diagram involving ratios $r_{1}$\ and $r_{2}$. The case $j = i$\ corresponds to a correct classification in the region of the diagram where the source is actually located, and the indexes $i$ and $j$ take the 4 values \hii, LINERs, TO, Seyferts). In other words,   $P_\mathrm{i,i}$\ with $i=1$\ is the probability that a source classified as  \hii, is really \hii, $P_\mathrm{i,j}$ with $i=2$, and $j=1$ is the probability that a source falling in the TO domain is misclassified, and that the correct classification is \hii.  Clearly,  $\sum_\mathrm{j} P_\mathrm{i,j} = 1, \forall i $.   $P_\mathrm{i,j}$\  can be written as follows:

\begin{equation}
P_\mathrm{i,j} = \int_{r_\mathrm{1,j,min}}^{r_\mathrm{1,j,max}}  
\left(\int_{r_\mathrm{2,j,min}(r_1)}^{r_\mathrm{2,j,max}(r_1)} \pi_2 (r_2) dr_2 \right)
\pi_1(r_1) dr_1
\label{eq:prob}
\end{equation}

\begin{eqnarray}
P_\mathrm{i,j} & = & \int_{r_\mathrm{1,j,min}}^{r_\mathrm{1,j,max}}  \left\{ \Pi_2(r_\mathrm{2,j,max} (r_1)) - \Pi_2(r_\mathrm{2,j,min}(r_1)) \right\}\\ 
&& \pi_1(r_1) dr_1\nonumber
\end{eqnarray}

We described here a DD as a plane with a dependence between the ratio on ordinate $r_2$ on the ratio  $r_1$ on abscissa,  $r_\mathrm{2,j,max} (r_1)$\ and $r_\mathrm{2,j,min} (r_1)$.  The probabilities are assigned assuming that errors on ratios  follow a log-normal distribution in the case of detection and define a probability density $\pi$ for each diagnostic ratio. $\Pi_2$ is the cumulative distribution integrated over $r_2$\ that is a function of $r_1$\ since the limits on $r_2$ are in general a function of $r_1$\ given by dividing lines of \citet{kewleyetal01} and \citet{kauffmannetal03}.    An underlying assumption is that the joint probability density $\pi$ can be factored as $\pi = \pi_1(r_1) \cdot \pi_2(r_2)$\ i.e., that the probabilities of the two diagnostic ratios are independent.  A similar approach was followed by \citet{manzerderobertis14}. 

The probability can be assigned also in case of upper and lower limits, considering the Kaplan-Meier (KM) estimator of the cumulative survival function \citep{feigelsonnelson85}. To keep our approach simple we approximate these cumulative distributions with an error  function and the probability density with   Gaussian distributions in the log-log plane for the diagnostic ratios \oiii/\hb, \nii/\ha, and \oii/\hb, along with  the probability density field associated with each KM estimator (dashed lines). The probability density is needed to compute the integral of Eq. \ref{eq:prob} in case a censored value appears on the ratio in abscissa.  If this approach is followed, it is possible to define a $P_\mathrm{ij}$\ for $i$=Seyfert, \hii, TO, and LINER.   Censored data analysis  is rigorous as long as line equivalent width and single intensity ratios are considered, but there is no established solution to the  statistical problem of univariate or 2D censoring with mixed censoring (i.e., upper and lower limit). 
The KM estimators    have been computed considering  detections + UL and  detections + LL separately.   The value of the probability estimates if upper (occurring most often) and lower limits (rarer) are included will be therefore heuristic. In addition, the K-M estimator of the survival function  depends on the observed distribution of data points which in turns depends on the instrumental capabilities of the survey, as well as on the intrinsic physical properties of the sample that is observed. Nonetheless, consideration of censored data provides a more realistic view of important aspects related to the prevalence of ELG  classes, and to the difference in distribution of parameters like line luminosity and equivalent width. Probabilities including upper limits were assigned  {  and are reported in the database table (Tab. \ref{tab:table}, P\_OII\_ \ldots, P\_NII\_ \ldots, etc. keys).}

\end{appendix}

\end{document}